\documentclass[journal,twoside]{IEEEtran}
\ifCLASSINFOpdf
\else
\fi

\usepackage{setspace}
\usepackage{graphicx}
\usepackage{amsmath}
\usepackage{epstopdf}
\usepackage{multirow}
\usepackage{color}
\usepackage{amsthm}
\usepackage{amssymb}
\usepackage{bigints}
\usepackage{bm}

\DeclareGraphicsExtensions{.pdf,.eps,.png,.jpg,.mps}
\DeclareMathOperator{\Tr}{Tr}
\DeclareMathOperator{\diag}{diag}
\DeclareMathOperator{\rank}{rank}

\newtheorem{theorem}{Theorem}[section]

\newtheorem{proposition}{Proposition}[section]

\newtheorem{lemma}{Lemma}[section]
\newcommand{\argmin}{\operatornamewithlimits{argmin}}
\newcommand\undermat[2]{%
  \makebox[0pt][l]{$\smash{\underbrace{\phantom{%
    \begin{matrix}#2\end{matrix}}}_{\text{$#1$}}}$}#2}
\makeatletter
\renewcommand*\env@matrix[1][\arraystretch]{%
  \edef\arraystretch{#1}%
  \hskip -\arraycolsep
  \let\@ifnextchar\new@ifnextchar
  \array{*\c@MaxMatrixCols c}}
\makeatother

\begin{document}
%
\title{Asynchronous Orthogonal Differential \\
Decoding for Multiple Access Channels}
%
%
%

\author{Sina~Poorkasmaei~and~Hamid~Jafarkhani,~\IEEEmembership{Fellow,~IEEE}
\thanks{This work was supported in part by the NSF under Award CCF-0963925. The material in this paper will be presented in part at the 2014 IEEE Global Communications Conference.}
\thanks{The authors are with the Center
for Pervasive Communications and Computing, University of California, Irvine,
CA 92697-2625 USA (e-mail: \{spoorkas, hamidj\}@uci.edu).}}

%
%

\markboth{IEEE Transactions on Wireless Communications}%
{Poorkasmaei and Jafarkhani: Asynchronous Orthogonal Differential Decoding for Multiple Access Channels}
%



\maketitle

\begin{abstract}
We propose several differential decoding schemes for asynchronous multi-user MIMO systems based on orthogonal space-time block codes (OSTBCs) where neither the transmitters nor the receiver has knowledge of the channel. First, we derive novel low complexity differential decoders by performing interference cancelation in time and employing different decoding methods. The decoding complexity of these schemes grows linearly with the number of users. We then present additional differential decoding schemes that perform significantly better than our low complexity decoders and outperform the existing synchronous differential schemes but require higher decoding complexity compared to our low complexity decoders. The proposed schemes work for any square OSTBC, any constant amplitude constellation, any number of users, and any number of receive antennas. Furthermore, we analyze the diversity of the proposed schemes and derive conditions under which our schemes provide full diversity. For the cases of two and four transmit antennas, we provide examples of PSK constellations to achieve full diversity. Simulation results show that our differential schemes provide good performance. To the best of our knowledge, the proposed differential detection schemes are the first differential schemes for asynchronous multi-user systems.
\end{abstract}


%
\IEEEpeerreviewmaketitle

\section{Introduction}
%
%
%
%

\IEEEPARstart{V}{arious} space-time modulation techniques to achieve transmit diversity have been proposed in the literature \cite{J05}. In most cases, it is assumed that the channel state information (CSI) is perfectly known at the receiver \cite{A98}, \cite{TJC99}. This is a reasonable assumption when the channel changes slowly and can be estimated by transmitting known training symbols. However, this is not always possible, and there is a tradeoff between frame length and accuracy of the channel estimation \cite{YDKP12}. Therefore, the effects of channel estimation error make it desirable to use schemes that avoid such an estimation.

Prior work has proposed many differential space-time coding schemes in which neither the transmitter nor the receiver knows the CSI. The first differential coding schemes based on orthogonal designs for multiple transmit antennas were proposed in \cite{TJ00} and \cite{JT01} with about 3-dB loss in performance compared to the corresponding coherent detection. Other examples of differential modulation schemes using space-time block codes (STBCs) and linear decoding complexity were proposed in \cite{GS02}-\cite{TC01}. A rate-one differential modulation scheme based on the quasi-orthogonal space-time block codes (QOSTBCs) \cite{J01} can be found in \cite{ZJ05}.

Multi-user detection schemes with simple coherent detection structures for multiple access channels (MACs) have garnered significant attention \cite{NSC98}-\cite{KJ08}. The main goal is to design a low complexity interference cancelation method for a MAC with $J$ users using only $J$ receive antennas. This is done for $N=2$ transmit antennas in \cite{NSC98} and for $J=2$ users in \cite{SAC01} using the properties of orthogonal space-time block codes (OSTBCs) \cite{TJC99}. To solve the problem for any number of users, any constellation, and any number of transmit antennas, \cite{KJ08} presents a method utilizing QOSTBCs with a moderate increase in decoding complexity. Space-time/frequency code design criteria for fading MIMO MACs and a code construction for two users have been derived in \cite{GB06}.

Differential modulation schemes for two-user MAC systems have been proposed in \cite{BH12}. These schemes have a high decoding complexity. In \cite{PJ13}, we proposed low complexity differential modulation schemes for two-user MIMO systems that achieve full transmit diversity. Moreover, we presented additional differential decoding schemes that provide full diversity, outperform the existing differential schemes, and work for any square OSTBC ($N \times N$ OSTBC).

All the existing multi-user differential schemes assume the transmission of the data by the users to be perfectly synchronized in time. To the best of our knowledge, a differential modulation scheme for asynchronous multi-user systems does not exist in the literature. In this paper, we design differential detection schemes for asynchronous multi-user MIMO systems where neither the transmitters nor the receiver knows the channel. Our main results are as follows:
\begin{enumerate}
  \item With a slow Rayleigh fading channel model for an asynchronous multi-user system, we present a differential encoder and derive novel low complexity differential decoders by performing interference cancelation in time and employing different decoding methods. The decoding complexity of these schemes grows linearly with the number of users.
  \item We also present additional differential decoding schemes that perform significantly better than our low complexity decoders and outperform the existing synchronous differential schemes, but need higher decoding complexity compared to our low complexity decoders.
  \item All the proposed decoders work for any square OSTBC, any constant amplitude constellation, any number of users, and any number of receive antennas.
  \item We analyze the diversity of our schemes and derive conditions under which the proposed schemes provide full diversity. For the cases of two and four transmit antennas, we provide examples of PSK constellations to achieve full diversity. Simulation results show that the proposed differential detection schemes provide good performance.
\end{enumerate}

The rest of the paper is organized as follows. In Section \ref{systemmodel}, we introduce the system model. In Section \ref{encoding}, we present the differential encoding for our asynchronous differential modulation schemes. The differential decoding schemes are put forward in Section \ref{decoding}. We analyze the diversity of our schemes in Section \ref{diversitygainanalysis}. Simulation results are provided in Section \ref{simulations}, and Section \ref{conclusions} concludes the paper.

\textbf{Notation:} We use boldface capital letters to denote matrices, boldface small letters to denote vectors, and super-scripts $(\cdot)^*$ and $(\cdot)^\dagger$ to denote conjugate and conjugate transpose, respectively. $\|\cdot\|_F$ indicates the Frobenius norm, and $E\left[\cdot\right]$ represents the expected value. Also, we use $\bm{I}_n$ and $\bm{0}_n$ to denote the $n \times n$ identity and zero matrices, respectively, and $\bm{0}_{m \times n}$ to denote the $m \times n$ zero matrix.

\section{System Model}
\label{systemmodel}

We consider a wireless communication system with $J$ users each with $N$ transmit antennas and one receiver with $M$ receive antennas with a quasi-static flat Rayleigh fading channel. We define $\bm{H}_j$, $j=1,\cdots,J$, as $M \times N$ channel fading matrices whose $(m,n)$th elements $h_{j,m,n}$ are the channel fading coefficients from transmit antenna $n$ to receive antenna $m$ for User $j$. The entries of $\bm{H}_j$, $j=1,\cdots,J$, are samples of independent zero-mean complex Gaussian random variables with a variance of 0.5 per real dimension.

In a practical set-up, the transmitters use pulse-shaping filters, and the receiver usually utilizes a matched filter to maximize the SNR. In such a scenario, the role of sampling is to provide a set of sufficient statistics for the detection of the received signals. Consider the $N \times 1$ signal vector transmitted by the $j$th transmitter
\begin{equation}
\bm{x}_j(t) = \sum_{k} \bm{s}_j(k) \psi(t-kT_s)
\label{eq0001}
\end{equation}
where $\bm{s}_j(\cdot)$ is the $N \times 1$ symbol vector, $T_s$ is a symbol duration, and $\psi(\cdot)$ is the pulse-shaping filter with a non-zero duration of at most $LT_s$ for some $L \in \mathbb{N}$ (i.e., $\psi(t)=0$, $|t|>\frac{L}{2}T_s$). We assume the average transmit power of each user is unity. The $M \times 1$ received signal vector is
\begin{equation}
\begin{array} {l@{}l}
\bm{y}(t) &= \displaystyle\sum_{j=1}^J \bm{H}_j \bm{x}_j(t-\tau_j) + \bm{n}(t) \\
&= \displaystyle\sum_{j=1}^J \bm{H}_j \displaystyle\sum_{k} \bm{s}_j(k) \psi(t-kT_s-\tau_j) + \bm{n}(t)
\end{array}
\label{eq0002}
\end{equation}
where $\bm{n}(t)$ is the $M \times 1$ complex white Gaussian noise vector, and the symbol vectors $\bm{s}_j(k)$ for the $j$th user are transmitted through the channel matrix $\bm{H}_j$ and received with a relative delay of $\tau_j$. We assume $\tau_j$ is fixed within a frame. Then, considering the transmission of a frame of $D$ symbol vectors $\bm{s}_j(1),\cdots,\bm{s}_j(D)$ and assuming $\bm{s}_j(k)=0$ for $k\notin\{1,\cdots,D\}$, the optimum maximum-likelihood (ML) receiver uses the log-likelihood cost function given by
\begin{equation}
\footnotesize{\begin{array} {l@{}l@{}l}
\bm{\Lambda} &=& \displaystyle\bigintsss \left\| \bm{y}(t)-\sum_{j=1}^J \bm{H}_j \sum_{k=1}^D \bm{s}_j(k) \psi(t-kT_s-\tau_j) \right\|_F^2 dt \\
&=& \displaystyle\bigintsss \left\| \bm{y}(t) \right\|_F^2 dt + \displaystyle\bigintsss \left\| \sum_{j=1}^J \bm{H}_j \sum_{k=1}^D \bm{s}_j(k) \psi(t-kT_s-\tau_j) \right\|_F^2 dt \\
&&- 2\operatorname{Re}\left\{\Tr\left[\displaystyle\sum_{j=1}^J\left(\displaystyle\sum_{k=1}^D \displaystyle\bigintsss \bm{y}(t) \psi^*(t-kT_s-\tau_j) dt \cdot \bm{s}_j^\dagger(k)\right) \cdot \bm{H}_j^\dagger\right]\right\}.
\end{array}}
\label{eq0003}
\end{equation}
Now, suppose that $\psi(\cdot)$, $T_s$ and $\tau_j$ are all known at the receiver and consider the RHS of the last equality in (\ref{eq0003}). The first integral depends only on $\bm{y}(t)$, which is the same for all possible information sequences, and thus can be ignored for ML decoding. Also, for a given sequence $\bm{s}_j(k)$, since all other quantities are known in coherent detection, the second integral can be calculated independent of the received signal. Finally, in terms of the received signal, it is sufficient to know only the last integral in order to perform ML decoding. Therefore, the output of the matched filter can be sampled at different sampling times associated with different transmitters to construct $\mathbf{y}_i(k)$ as follows
\begin{equation}
\small{\begin{array} {l@{}l}
\mathbf{y}_i(k) = \displaystyle\bigintsss_{(k-\frac{L}{2})T_s+\tau_i}^{(k+\frac{L}{2})T_s+\tau_i} \bm{y}(t) \psi^*(t-kT_s-\tau_i) dt, &
\begin{array} {c}
i=1,\cdots,J, \\
k=1,\cdots,D.
\end{array}
\end{array}}
\label{eq0004}
\end{equation}
Clearly, the operations in (\ref{eq0004}) do not destroy any information that is valuable in deciding which symbols were transmitted, and thus these samples constitute a set of sufficient statistics for detecting all symbols. To simplify the notation, we assume that $\tau_1=0$, $\tau_1<\tau_2<\cdots<\tau_J<T_s$, and $\tau_{(i_1 + i_2 \cdot J)}=\tau_{i_1} + i_2 \cdot T_s$ ($\forall \ i_1,i_2 \in \mathbb{Z}$). We can write each integral in (\ref{eq0004}) as the sum of multiple integrals on smaller intervals. Then, we can scale the resulting integrals for simplification in notation and construct a new set consisting of all these integrals to obtain another set of sufficient statistics for detection of all symbols as
\begin{equation}
\small{\begin{array} {l@{}l}
\bm{y}_i(d) &= \frac{T_s}{\tau_{i+1,i}}\displaystyle\bigintsss_{(d-\frac{L}{2})T_s+\tau_i}^{(d-\frac{L}{2})T_s+\tau_{i+1}} \bm{y}(t) \psi^*(t-dT_s-\tau_i) dt \\
&= \displaystyle\sum_{j=1}^J \bm{H}_j \displaystyle\sum_{r=0}^L \bm{s}_j(d-r)\alpha_{j,i}(r)+\bm{n}_i(d), \\
&\qquad \qquad \qquad \quad \ i=1,\cdots,J, \quad d=1,\cdots,D+L,
\end{array}}
\label{eq0005}
\end{equation}
where $\tau_{i_1,i_2} = \tau_{i_1}-\tau_{i_2}$, $\forall \ i_1,i_2$,
\begin{equation}
\small{\begin{array} {l@{}l}
\bm{n}_i(d) &= \frac{T_s}{\tau_{i+1,i}}\displaystyle\bigintsss_{(d-\frac{L}{2})T_s+\tau_i}^{(d-\frac{L}{2})T_s+\tau_{i+1}} \bm{n}(t) \psi^*(t-dT_s-\tau_i) dt, \\
\alpha_{j,i}(r) &= \frac{T_s}{\tau_{i+1,i}}\displaystyle\bigintsss_{(d-\frac{L}{2})T_s+\tau_i}^{(d-\frac{L}{2})T_s+\tau_{i+1}} \begin{array} {l}
\psi(t-(d-r)T_s-\tau_j) \\
\qquad \quad \ \cdot \psi^*(t-dT_s-\tau_i) dt
\end{array} \\
&= \frac{T_s}{\tau_{i+1,i}}\displaystyle\bigintsss_{-\frac{L}{2}T_s}^{\tau_{i+1,i}-\frac{L}{2}T_s} \psi(t+rT_s-\tau_{j,i}) \psi^*(t) dt.
\end{array}}
\label{eq0006}
\end{equation}
Note that the last element of the set, $\bm{y}_J(D+L)$, is not obtained by splitting and scaling the integrals in (\ref{eq0004}). However, we make the notation simpler by adding it to the set, and the result is still a set of sufficient statistics. Also, notice that $\alpha_{j,i}(r)=0$ for $r\notin\{0,\cdots,L\}$. Therefore, the index $r$ in (\ref{eq0005}) and (\ref{eq0006}) ranges from $0$ to $L$. Moreover, $\bm{n}_i(d)$, $\forall \ i,d$, are independent zero-mean complex Gaussian random vectors with covariance matrices $E\left[\bm{n}_i(d)\bm{n}_i^\dagger(d)\right] = \frac{(\textrm{SNR})^{-1} T_s \alpha_{i,i}(0)}{\tau_{i+1,i}} \cdot \bm{I}_M$ where SNR is the ratio of the average transmit power to the noise power. Let
\begin{equation}
\begin{array} {c}
\bm{Y}(d) = \left(\bm{y}_1(d),\cdots,\bm{y}_J(d)\right), \ \bm{N}(d) = \left(\bm{n}_1(d),\cdots,\bm{n}_J(d)\right), \\
\bm{\alpha}_j(r) = \left(\alpha_{j,1}(r),\cdots,\alpha_{j,J}(r)\right).
\end{array}
\label{eq0007}
\end{equation}
Then, the received samples can be written in a matrix form as
\begin{equation}
\bm{Y} = \sum_{j=1}^J \bm{H}_j \bm{S}_j \bm{A}_j + \bm{N}
\label{eq0008}
\end{equation}
where $\bm{Y} = \left(\bm{Y}(1),\cdots,\bm{Y}(D+L)\right)$, $\bm{S}_j = \left(\bm{s}_j(1),\cdots,\right.$ $\left.\bm{s}_j(D)\right)$, $\bm{N} = \left(\bm{N}(1),\cdots,\bm{N}(D+L)\right)$ are $M \times (D+L)J$, $N \times D$ and $M \times (D+L)J$ matrices, respectively, and $\bm{A}_j$ is a $D \times (D+L)J$ matrix given by
\begin{equation}
\scriptsize{\setlength\arraycolsep{0.75pt}
\bm{A}_j = \begin{pmatrix}
               \bm{\alpha}_j(0) & \bm{\alpha}_j(1) & \cdots & \bm{\alpha}_j(L) & \bm{0}_{1 \times J} & \bm{0}_{1 \times J} & \cdots & \bm{0}_{1 \times J} & \cdots & \bm{0}_{1 \times J} \\
               \bm{0}_{1 \times J} & \bm{\alpha}_j(0) & \bm{\alpha}_j(1) & \cdots & \bm{\alpha}_j(L) & \bm{0}_{1 \times J} & \cdots & \bm{0}_{1 \times J} & \cdots & \bm{0}_{1 \times J} \\
               \ddots & \ddots & \ddots & \ddots & \ddots & \ddots & \ddots & \ddots & \ddots & \ddots \\
               \bm{0}_{1 \times J} & \cdots & \bm{0}_{1 \times J} & \cdots & \bm{0}_{1 \times J} & \bm{\alpha}_j(0) & \bm{\alpha}_j(1) & \cdots & \bm{\alpha}_j(L) & \bm{0}_{1 \times J} \\
               \bm{0}_{1 \times J} & \cdots & \bm{0}_{1 \times J} & \cdots & \bm{0}_{1 \times J} & \bm{0}_{1 \times J} & \bm{\alpha}_j(0) & \bm{\alpha}_j(1) & \cdots & \bm{\alpha}_j(L) \\
           \end{pmatrix}.}
\label{eq0009}
\end{equation}
For the sake of simplicity, in this paper we consider the case where $L=1$ and the pulse-shaping filter is a rectangular pulse
\begin{equation}
\psi(t) = \left\{ \begin{array}{ll}
1/\sqrt{T_s} &, -T_s/2\le t<T_s/2\\
0 &, \textrm{otherwise}
\end{array} \right..
\label{eq0010}
\end{equation}
Then, it can be easily seen from (\ref{eq0006}) that
\begin{equation}
\alpha_{j,i}(0) = \left\{ \begin{array}{ll}
1 &, j\le i\\
0 &, \textrm{otherwise}
\end{array} \right.,\ \
\alpha_{j,i}(1) = \left\{ \begin{array}{ll}
1 &, j>i\\
0 &, \textrm{otherwise}
\end{array} \right..
\label{eq0011}
\end{equation}
Therefore, in this case, using (\ref{eq0007}), (\ref{eq0009}) and (\ref{eq0011}), $\bm{A}_j$ becomes
\begin{equation}
\small{\setlength\arraycolsep{3pt}
\bm{A}_j = \begin{pmatrix}[1.9]
               \smash{\overbrace{0 \cdots 0}^{j-1 \textrm{times}}} & \smash{\overbrace{1 \cdots 1}^{J\ \textrm{times}}} & 0 \cdots 0 & \cdots & 0 \cdots 0 & 0 \cdots 0 \\
               0 \cdots 0 & 0 \cdots 0 & \smash{\overbrace{1 \cdots 1}^{J\ \textrm{times}}} & \cdots & 0 \cdots 0 & 0 \cdots 0 \\
               \ddots & \ddots & \ddots & \ddots & \ddots & \ddots \\
               0 \cdots 0 & 0 \cdots 0 & 0 \cdots 0 & \cdots & \smash{\overbrace{1 \cdots 1}^{J\ \textrm{times}}} & \smash{\overbrace{0 \cdots 0}^{J-j+1 \textrm{times}}} \\
           \end{pmatrix},}
\label{eq0012}
\end{equation}
and $\bm{n}_i(d)$, $\forall \ i,d$, become independent zero-mean complex Gaussian random vectors with covariance matrices $E\left[\bm{n}_i(d)\bm{n}_i^\dagger(d)\right] = \frac{(\textrm{SNR})^{-1} T_s}{\tau_{i+1,i}} \cdot \bm{I}_M$. Note that we have not made any assumption about the values of delay differences. However, because of the scaling factor of $\frac{T_s}{\tau_{i+1,i}}$ used in (\ref{eq0005}), $A_j$ includes only 0s and 1s. Therefore, the values of $\tau_1, \tau_2, \cdots, \tau_J$ only appear in the noise covariance matrices in our system model.

In what follows, we consider the received signals in size $TJ$ blocks of $(\bm{y}_1(Tl+1), \cdots, \bm{y}_J(Tl+1), \cdots, \bm{y}_1(Tl+T), \cdots, \bm{y}_J(Tl+T))$, for $l=0,1,\cdots$, and with a small abuse of the notation, we denote them as $(\bm{y}_{1,1}^l, \cdots, \bm{y}_{1,J}^l, \cdots,\bm{y}_{T,1}^l, \cdots, \bm{y}_{T,J}^l)$. Similarly, we denote the noise terms $(\bm{n}_1(Tl+1), \cdots, \bm{n}_J(Tl+1), \cdots, \bm{n}_1(Tl+T), \cdots, \bm{n}_J(Tl+T))$ as $(\bm{n}_{1,1}^l, \cdots, \bm{n}_{1,J}^l, \cdots, \bm{n}_{T,1}^l, \cdots, \bm{n}_{T,J}^l)$, for $l=0,1,\cdots$. We define $K$ as the number of data symbols transmitted during one block. The channels are assumed to be unknown at both the transmitters and the receiver.

\section{Differential Encoding}
\label{encoding}

\begin{figure}
\begin{center}
  \includegraphics[scale=0.23]{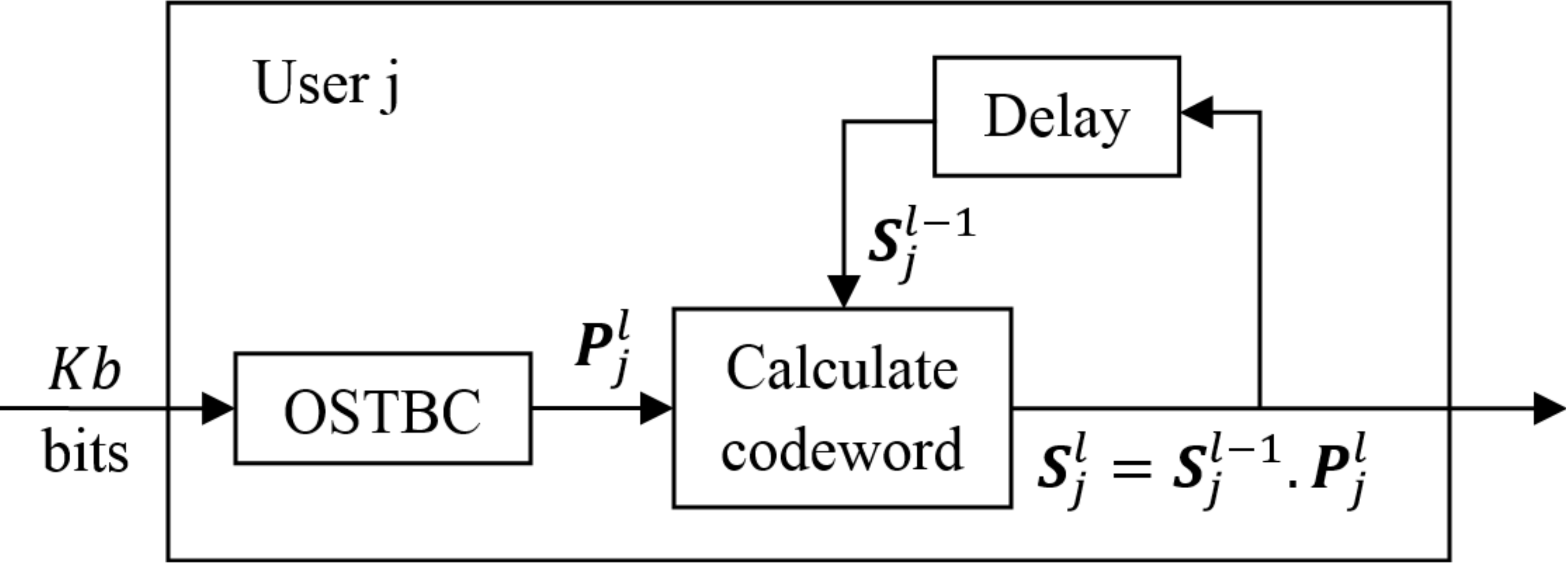}
  \caption{Block diagram of differential encoder.\label{Fig1}}
\end{center}
\end{figure}

In this section, we describe our differential encoding scheme for User $j=1,\cdots,J$. The block diagram of the differential encoder is the same as that of a synchronized system and is shown in Fig. \ref{Fig1}.
The main difference with the synchronous case \cite{BH12}, \cite{PJ13} is that different users do not need to employ different constellations. At a transmission rate of $b$ bits/(s \nolinebreak Hz), we use a constant amplitude signal constellation with $2^b$ elements such as $2^b$-PSK with an appropriate normalization to make the transmitted codewords unitary. Similar to the case of a single user, extension to other constellations is possible. For each block of $Kb$ bits, User $j$ selects $K$ symbols and transmits them using an $N \times N$ OSTBC. This transmitted codeword also depends on the codeword and symbols transmitted in the previous block. We assume the input bits are the outputs of independent uniformly distributed random variables.

The encoding starts with the transmission of arbitrary $N \times N$ OSTBCs $\bm{S}_j^0$ and $\bm{S}_j^1$. As in the case of a single user, we could transmit only one OSTBC instead of two and the system would still work with minor changes. For block $l$, we use the $Kb$ input bits to pick $K$ symbols $p_{j,1}^l, \cdots, p_{j,K}^l$ from the signal constellation and construct the corresponding square OSTBC, $\bm{P}_j^l$.
Assuming that $\bm{S}_j^{l-1}$ is the codeword of User $j$ for the $(l-1)$th block, we calculate $\bm{S}_j^l$ by
\begin{equation}
\bm{S}_j^l=\bm{S}_j^{l-1} \cdot \bm{P}_j^l
\label{eq0013}
\end{equation}
and then transmit it at block $l$. Note that the generated codeword $\bm{S}_j^l$ will be orthogonal as well. Later, in Section \ref{diversitygainanalysis}, we analyze the diversity of the proposed schemes and derive conditions under which our schemes provide full diversity.

\section{Differential Decoding}
\label{decoding}

In this section, we present differential decoding schemes for all users. First, we derive novel low complexity decoders by performing interference cancelation in time and employing different decoding methods. The decoding complexity of these decoders increases linearly with the number of users. We then present additional decoding schemes that perform significantly better compared to our low complexity decoders and outperform the existing synchronous differential schemes. All the proposed decoders work for any square OSTBC, any constant amplitude constellation, any number of users, and any number of receive antennas. We assume that the channel is unchanged within three consecutive time blocks.\footnotemark\footnotetext[1]{As will become clear later, the channel could be assumed to be unchanged within a shorter period of time and our schemes would still work with minor changes.}

\subsection{Low Complexity Decoding Schemes}
\label{lcdecoding}

In this subsection, we introduce low complexity decoders for $J$ users with $N$ transmit antennas through several decoding methods. First, we start with a simple example for $J=2$ users and $N=2$ transmit antennas to illustrate the main ideas behind our low complexity decoders. In what follows, we describe the decoding procedure for User 2. We use a subscript 2 for the quantities used in decoding the signals of User 2 to distinguish them from those of User 1.

Note that the input-output relationship in (\ref{eq0008}) contains the signals for the entire frame. We can rewrite (\ref{eq0008}) for a single time block $l>0$ as
\begin{equation}
\footnotesize{\begin{array}{l}
\begin{pmatrix}
    \bm{y}_{1,1}^l, \bm{y}_{1,2}^l, \bm{y}_{2,1}^l, \bm{y}_{2,2}^l
\end{pmatrix} = \bm{H}_1 \begin{pmatrix}
    \bm{S}_1^{l-1}, \bm{S}_1^l
\end{pmatrix}\begin{pmatrix}[0.5]
               0 & 0 & 0 & 0 \\
               0 & 0 & 0 & 0 \\
               1 & 1 & 0 & 0 \\
               0 & 0 & 1 & 1
           \end{pmatrix} \\
           \qquad \quad + \bm{H}_2 \begin{pmatrix}
    \bm{S}_2^{l-1}, \bm{S}_2^l
\end{pmatrix}\begin{pmatrix}[0.5]
               0 & 0 & 0 & 0 \\
               1 & 0 & 0 & 0 \\
               0 & 1 & 1 & 0 \\
               0 & 0 & 0 & 1
           \end{pmatrix} + \begin{pmatrix}
    \bm{n}_{1,1}^l, \bm{n}_{1,2}^l, \bm{n}_{2,1}^l, \bm{n}_{2,2}^l
\end{pmatrix}.
\end{array}}
\label{eq0514}
\end{equation}
Then, note that the interference of User 1 on User 2 can be canceled by subtracting $\bm{y}_{t,1}^l$ from $\bm{y}_{t,2}^l$ for $t=1,2$ as follows
\begin{equation}
\footnotesize{\begin{pmatrix}\bm{\bar{y}}_{1,2}^l, \bm{\bar{y}}_{2,2}^l\end{pmatrix} =
\bm{H}_2 \begin{pmatrix}
    \bm{S}_2^{l-1}, \bm{S}_2^l
\end{pmatrix}\begin{pmatrix}[0.5]
               0 & 0 \\
               -1 & 0 \\
               1 & -1 \\
               0 & 1
           \end{pmatrix}
+ \begin{pmatrix}\bm{\bar{n}}_{1,2}^l, \bm{\bar{n}}_{2,2}^l\end{pmatrix}}
\label{eq0516}
\end{equation}
where $\bm{\bar{y}}_{t,2}^l = \bm{y}_{t,2}^l-\bm{y}_{t,1}^l, \bm{\bar{n}}_{t,2}^l = \bm{n}_{t,2}^l-\bm{n}_{t,1}^l$ for $t=1,2$. Considering (\ref{eq0516}) for more consecutive time slots and using simple algebra, one may obtain
\begin{equation}
\footnotesize{\begin{array}{l}
\underbrace{\left(
    \bm{\bar{y}}_{2,2}^{l-2}, \bm{\bar{y}}_{1,2}^{l-1}, \bm{\bar{y}}_{2,2}^{l-1}, \bm{\bar{y}}_{1,2}^l, \bm{\bar{y}}_{2,2}^l
\right)}_{\bm{\bar{Y}}_2^l} \\
\qquad = \bm{H}_2\cdot\begin{pmatrix}
    \bm{S}_2^{l-2}, \bm{S}_2^{l-1}, \bm{S}_2^l \\
\end{pmatrix}\cdot\bm{\bar{A}} + \underbrace{\left(
    \bm{\bar{n}}_{2,2}^{l-2}, \bm{\bar{n}}_{1,2}^{l-1}, \bm{\bar{n}}_{2,2}^{l-1}, \bm{\bar{n}}_{1,2}^l, \bm{\bar{n}}_{2,2}^l
\right)}_{\bm{\bar{N}}_2^l} \\
\qquad = \bm{H}_2 \ \bm{S}_2^{l-2} \ \underbrace{\left(
    \bm{I}_2, \bm{P}_2^{l-1}, \bm{P}_2^{l-1}\bm{P}_2^l
\right)}_{\bm{U}_2^l} \ \bm{\bar{A}} + \bm{\bar{N}}_2^l
\end{array}}
\label{eq0518}
\end{equation}
where
\begin{equation}
\small{\bm{\bar{A}} = \begin{pmatrix}[0.75]
               -1 & 0 & 0 & 0 & 0 \\
               1 & -1 & 0 & 0 & 0 \\
               0 & 1 & -1 & 0 & 0 \\
               0 & 0 & 1 & -1 & 0 \\
               0 & 0 & 0 & 1 & -1 \\
               0 & 0 & 0 & 0 & 1
           \end{pmatrix}.}
\label{eq0519}
\end{equation}
Now, to obtain our low complexity decoders, we note that when conditioned on $\bm{P}_2^{l-1}, \bm{P}_2^l$, the matrix $\bm{\bar{Y}}_2^l$ is Gaussian with conditional probability density function (pdf)
\begin{equation}
P\left(\bm{\bar{Y}}_2^l \left| \bm{P}_2^{l-1}, \bm{P}_2^l\right.\right) \propto \frac{\exp\left\{-\Tr\left[\bm{\bar{Y}}_2^l\cdot(\bm{\bar{V}}_2^l)^{-1}\cdot(\bm{\bar{Y}}_2^l)^\dagger\right]\right\}}{\left[\det(\bm{\bar{V}}_2^l)\right]^M}
\label{eq0016}
\end{equation}
where $\bm{\bar{V}}_2^l$ is the covariance matrix given by $\bm{\bar{V}}_2^l = (\bm{U}_2^l \ \bm{\bar{A}})^\dagger \cdot (\bm{U}_2^l \ \bm{\bar{A}}) + (\textrm{SNR})^{-1} T_s (\tau_{3,2}^{-1}+\tau_{2,1}^{-1}) \cdot \bm{I}_5$. Therefore, we can define our first low complexity decoder as
\begin{equation}
\left \{ \bm{\hat{P}}_2^{l-1}, \bm{\hat{P}}_2^l \right \} = \displaystyle \argmin_{\bm{P}_2^{l-1}, \bm{P}_2^l} \bm{\Lambda}_2^l \left( \bm{P}_2^{l-1},\bm{P}_2^l \right)
\label{eq0017}
\end{equation}
where $\bm{\Lambda}_2^l \left( \bm{P}_2^{l-1},\bm{P}_2^l \right)$ is given by
\begin{equation}
\bm{\Lambda}_2^l \left( \bm{P}_2^{l-1},\bm{P}_2^l \right) = M\cdot\ln\left[\det(\bm{\bar{V}}_2^l)\right] + \Tr\left[\bm{\bar{Y}}_2^l\cdot(\bm{\bar{V}}_2^l)^{-1}\cdot(\bm{\bar{Y}}_2^l)^\dagger\right].
\label{eq0018}
\end{equation}

We now consider the general case of $J$ users with $N$ transmit antennas and present our low complexity decoders. We illustrate the decoding process for User $j=1,\cdots,J$. In Method 0, we derive a low complexity decoder by canceling the interference of all users on User $j$ and then performing ML decoding. Based on the decoder in Method 0, we then use Methods 1 and 2 presented in \cite{PJ13} to improve the performance. These methods use dynamic programming (DP) to efficiently decode the transmitted data signals. As we will see later, the tradeoff for better performance of our differential schemes using Method 2 compared to that of Method 1 is the decoding delay (i.e., the number of time blocks it takes until the transmitted signals at a given time block are decoded by the receiver). Finally, using the decoder in Method 0, we present another decoding method (Method 3) to further reduce the decoding complexity while maintaining good performance.


\textit{Method 0:} We use the following proposition to design our low complexity decoders:
\begin{proposition} For any $l \geq 2$, the following relationship holds between the received signals and the transmitted signals of User $j=1,\cdots,J$
\begin{equation}
\bm{\bar{Y}}_j^l = \bm{H}_j \ \bm{S}_j^{l-2} \ \bm{U}_j^l \ \bm{\bar{A}} + \bm{\bar{N}}_j^l
\label{eq0014}
\end{equation}
where $\bm{\bar{A}}$ is a $3T \times 3T-1$ matrix given by
\begin{equation}
\small{\begin{array} {c}
\bm{\bar{A}} = \begin{pmatrix}[0.75]
               -1 & 0 & 0 & \cdots & 0 & 0 \\
               1 & -1 & 0 & \cdots & 0 & 0 \\
               0 & 1 & -1 & \cdots & 0 & 0 \\
               0 & 0 & 1 & \cdots & 0 & 0 \\
               \ddots & \ddots & \ddots & \ddots & \ddots & \ddots \\
               0 & 0 & 0 & \cdots & -1 & 0 \\
               0 & 0 & 0 & \cdots & 1 & -1 \\
               0 & 0 & 0 & \cdots & 0 & 1
           \end{pmatrix}, \\\\
\bm{\bar{Y}}_j^l = \left(
    \bm{\bar{y}}_{2,j}^{l-2}, \cdots, \bm{\bar{y}}_{T,j}^{l-2}, \bm{\bar{y}}_{1,j}^{l-1}, \cdots, \bm{\bar{y}}_{T,j}^{l-1}, \bm{\bar{y}}_{1,j}^l, \cdots, \bm{\bar{y}}_{T,j}^l
\right), \\
\bm{\bar{N}}_j^l = \left(
    \bm{\bar{n}}_{2,j}^{l-2}, \cdots, \bm{\bar{n}}_{T,j}^{l-2}, \bm{\bar{n}}_{1,j}^{l-1}, \cdots, \bm{\bar{n}}_{T,j}^{l-1}, \bm{\bar{n}}_{1,j}^l, \cdots, \bm{\bar{n}}_{T,j}^l
\right), \\
\bm{U}_j^l = \left(
    \bm{I}_N, \bm{P}_j^{l-1}, \bm{P}_j^{l-1}\bm{P}_j^l
\right),
\end{array}}
\label{eq0015}
\end{equation}
where $\bm{\bar{y}}_{t,j}^l = \bm{y}_{t,j}^l-\bm{y}_{t,j-1}^l, \bm{\bar{n}}_{t,j}^l = \bm{n}_{t,j}^l-\bm{n}_{t,j-1}^l$ for $t=1,\cdots,T$ and $\forall \ l$ (assuming that $\bm{y}_{t,0}^l, \bm{n}_{t,0}^l$, respectively, denote $\bm{y}_{t-1,J}^l, \bm{n}_{t-1,J}^l$ if $t\neq1$, and $\bm{y}_{T,J}^{l-1}, \bm{n}_{T,J}^{l-1}$ if $t=1$).
\begin{proof}
See Appendix A.
\end{proof}
\end{proposition}

Equation (\ref{eq0014}) is the main property used to design our low complexity differential decoding algorithm, where the interference of all users on User $j$ is completely canceled. Therefore, it can be utilized to decode the transmitted signals without interference. Notice that $\bm{\bar{Y}}_j^l$ starts from $\bm{\bar{y}}_{2,j}^{l-2}$ instead of $\bm{\bar{y}}_{1,j}^{l-2}$. We could consider using $\bm{\bar{y}}_{1,j}^{l-2}$ and other previously received signals to improve the performance of our scheme. However, that would cause additional inter-block interference from the previously transmitted signals of User $j$, which would then increase the decoding complexity. It is easy to see from (\ref{eq0014}) that when conditioned on $\bm{P}_j^{l-1}, \bm{P}_j^l$, the matrix $\bm{\bar{Y}}_j^l$ is Gaussian with conditional pdf
\begin{equation}
P\left(\bm{\bar{Y}}_j^l \left| \bm{P}_j^{l-1}, \bm{P}_j^l\right.\right) \propto \frac{\exp\left\{-\Tr\left[\bm{\bar{Y}}_j^l\cdot(\bm{\bar{V}}_j^l)^{-1}\cdot(\bm{\bar{Y}}_j^l)^\dagger\right]\right\}}{\left[\det(\bm{\bar{V}}_j^l)\right]^M}
\label{eq0016}
\end{equation}
where $\bm{\bar{V}}_j^l$ is the covariance matrix given by $\bm{\bar{V}}_j^l = (\bm{U}_j^l \ \bm{\bar{A}})^\dagger \cdot (\bm{U}_j^l \ \bm{\bar{A}}) + (\textrm{SNR})^{-1} T_s (\tau_{j+1,j}^{-1}+\tau_{j,j-1}^{-1}) \cdot \bm{I}_{3T-1}$. We are now prepared to present our first low complexity differential decoding scheme. One approach is to decode $\bm{P}_j^{l-1}$ and $\bm{P}_j^l$ jointly based on (\ref{eq0016}). Therefore, we define the Inter-Time Interference Cancelation (ITIC) decoding using Method 0 as
\begin{equation}
\left \{ \bm{\hat{P}}_j^{l-1}, \bm{\hat{P}}_j^l \right \} = \displaystyle \argmin_{\bm{P}_j^{l-1}, \bm{P}_j^l} \bm{\Lambda}_j^l \left( \bm{P}_j^{l-1},\bm{P}_j^l \right)
\label{eq0017}
\end{equation}
where $\bm{\Lambda}_j^l \left( \bm{P}_j^{l-1},\bm{P}_j^l \right)$ is given by
\begin{equation}
\bm{\Lambda}_j^l \left( \bm{P}_j^{l-1},\bm{P}_j^l \right) = M\cdot\ln\left[\det(\bm{\bar{V}}_j^l)\right] + \Tr\left[\bm{\bar{Y}}_j^l\cdot(\bm{\bar{V}}_j^l)^{-1}\cdot(\bm{\bar{Y}}_j^l)^\dagger\right].
\label{eq0018}
\end{equation}
Notice that for $l = 2$ in $\bm{U}_j^l$, $\bm{P}_j^1 = (\bm{S}_j^0)^\dagger \bm{S}_j^1$ is the arbitrary data matrix at block 1 and is known at both the encoder and decoder. When using this scheme, information provided by (\ref{eq0016}) at time blocks other than $l$ is ignored, and thus some performance is lost. To avoid such losses, we also propose additional decoding schemes using Methods 1 and 2 presented in \cite{PJ13} to efficiently decode the signals transmitted by the users. Note that we use the cost function of the ITIC decoder using Method 0 as described above, and thus the corresponding decoders using Methods 1 and 2 as presented in this paper are different from the decoders presented in \cite{PJ13}. In what follows, we summarize the description of the ITIC decoders using Methods 1 and 2 based on the cost function of the ITIC decoder using Method 0. We refer the interested reader to \cite{PJ13} for the details on derivations.

\begin{figure}
\begin{center}
  \includegraphics[scale=0.23]{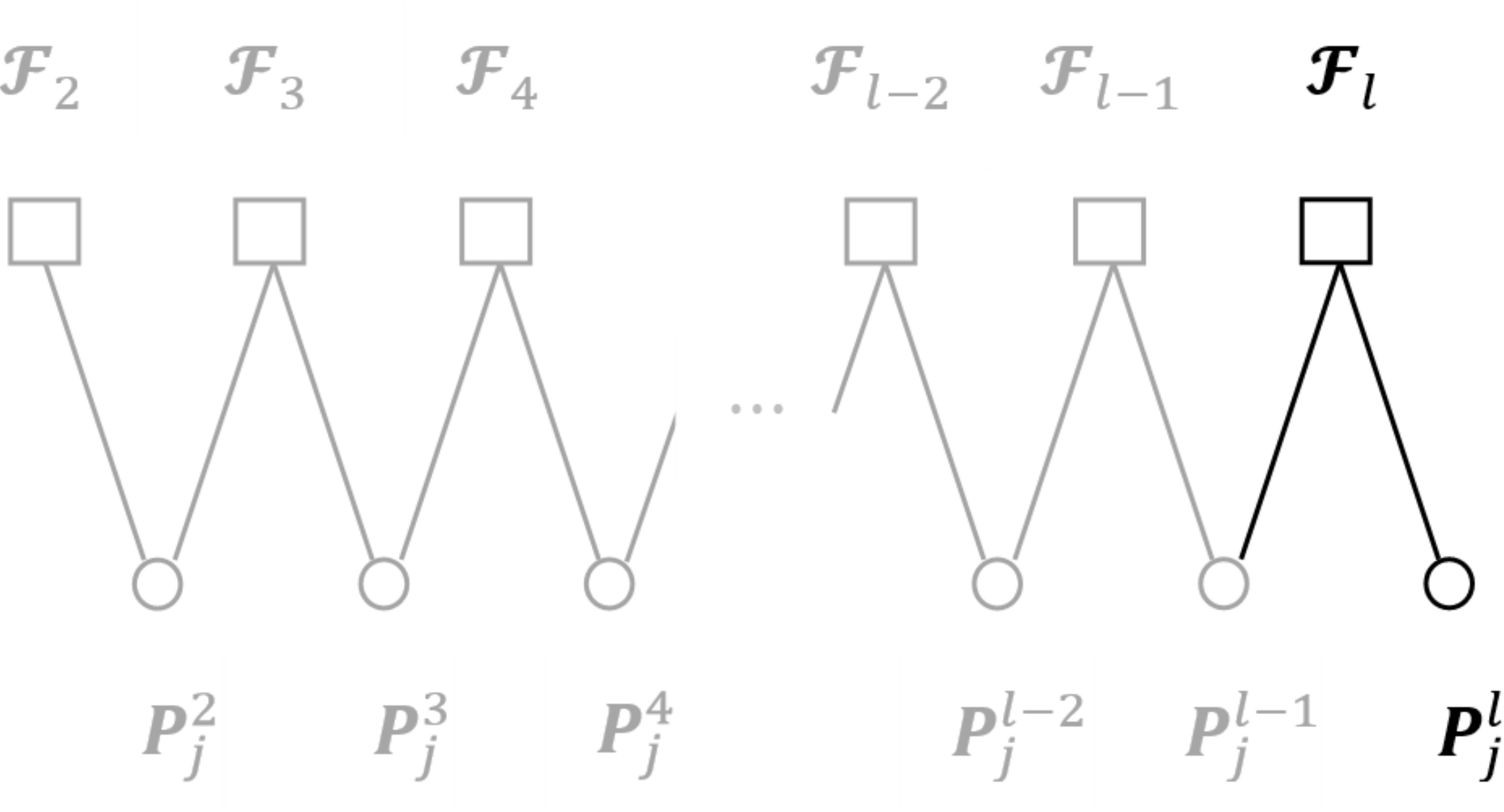}
  \caption{Chain corresponding to the decoding of $\bm{P}_j^{l}$.\label{Fig2}}
\end{center}
\end{figure}

\textit{Method 1 (Causal DP):} In Method 1, we decode $\bm{P}_j^{l}$ based on (\ref{eq0016}) for all blocks $\ell=2, \cdots ,l$ together. We utilize DP to efficiently find the best possible data matrix that maximizes an approximation for the conditional pdf of $\bm{\bar{Y}}_j^2, \cdots, \bm{\bar{Y}}_j^l$ given the data matrices $\bm{P}_j^2, \cdots ,\bm{P}_j^{l}$. Using (\ref{eq0016}) and ignoring the correlations of $\bm{\bar{Y}}_j^\ell$ at different blocks $\ell=2, \cdots ,l$ given the data matrices, we consider the following:
\begin{equation}
\begin{array}{l@{}l}
f_1 \left( \bm{P}_j^2, \cdots ,\bm{P}_j^{l} \right) &\propto \displaystyle\prod_{\ell=2}^{l} \exp \left\{ -\bm{\Lambda}_j^\ell \left( \bm{P}_j^{\ell-1},\bm{P}_j^\ell \right) \right\} \\
&= \exp \left\{ -\displaystyle\sum_{\ell=2}^{l} \bm{\Lambda}_j^\ell \left( \bm{P}_j^{\ell-1},\bm{P}_j^\ell \right) \right\}.
\end{array}
\label{eq116}
\end{equation}
In order to maximize the above function, we only need to minimize $\sum_{\ell=2}^{l} \bm{\Lambda}_j^\ell \left( \bm{P}_j^{\ell-1},\bm{P}_j^\ell \right)$. For any block $l \geq 2$, we define the ITIC decoding using Method 1 as
\begin{equation}
\bm{\hat{P}}_j^l = \displaystyle \argmin_{\bm{P}_j^l} \bm{\Phi}_j^l \left( \bm{P}_j^l \right)
\label{eq17}
\end{equation}
where $\bm{\Phi}_j^l \left( \bm{P}_j^l \right)$ is defined as
\begin{equation}
\small{\bm{\Phi}_j^l \left( \bm{P}_j^l \right) \triangleq \left\{ \begin{array}{ll}
\bm{\Lambda}_j^2 \left( \bm{P}_j^1,\bm{P}_j^2 \right) &, l=2\\
\displaystyle \min_{\bm{P}_j^2, \cdots ,\bm{P}_j^{l-1}} \sum_{\ell=2}^l \bm{\Lambda}_j^\ell \left( \bm{P}_j^{\ell-1},\bm{P}_j^\ell \right) &, \textrm{otherwise}
\end{array} \right..}
\label{eq18}
\end{equation}
The optimization problem in (\ref{eq18}) can be efficiently solved by utilizing DP. Using (\ref{eq18}), it is easy to show that for $l > 2$, we have
\begin{equation}
\bm{\Phi}_j^l \left( \bm{P}_j^l \right) = \displaystyle \min_{\bm{P}_j^{l-1}} \left \{ \bm{\Phi}_j^{l-1} \left( \bm{P}_j^{l-1} \right) + \bm{\Lambda}_j^l \left( \bm{P}_j^{l-1},\bm{P}_j^l \right) \right \}.
\label{eq19}
\end{equation}
As a result of storing the cost function of the previous block, $\mathbf{\Phi}_j^{l-1} \left( \bm{P}_j^{l-1} \right)$, we only need to perform an optimization over $\bm{P}_j^{l-1}$ for each possible data matrix $\bm{P}_j^l$ at time block $l$. That is, for each possible data matrix $\bm{P}_j^l$, in lieu of solving the optimization problem in (\ref{eq18}) over all data matrices for the previous blocks, $\bm{P}_j^2, \cdots ,\bm{P}_j^{l-1}$, we can solve the optimization problem in (\ref{eq19}) over the data matrix of only one block, $\bm{P}_j^{l-1}$, as illustrated in Fig. \ref{Fig2}.
The optimization in (\ref{eq19}) corresponds to the black path, while the optimization for the previous blocks corresponds to the gray path.

\textit{Method 2 (Non-Causal DP):} In Method 2, we consider some non-overlapping windows of blocks and decode the transmitted symbols within each window together. Note that since the decoding of each block may depend on future blocks in the same window, this method will cause some additional delay. However, since more information is used, the performance will improve as well.

Using Method 2, in the $m$th stage of decoding, $m \geq 1$, we decode the data matrices at blocks $k_{m-1}+1, \cdots, k_m$ where $k_0=1$ and $k_0 < k_1 < k_2 < \cdots$. We consider the following:
\begin{equation}
\begin{array}{l@{}l}
f_2 \left( \bm{P}_j^2, \cdots ,\bm{P}_j^{k_m} \right) &\propto \displaystyle\prod_{\ell=2}^{k_m} \exp \left\{ -\bm{\Lambda}_j^\ell \left( \bm{P}_j^{\ell-1},\bm{P}_j^\ell \right) \right\} \\
&= \exp \left\{ -\displaystyle\sum_{\ell=2}^{k_m} \bm{\Lambda}_j^\ell \left( \bm{P}_j^{\ell-1},\bm{P}_j^\ell \right) \right\}.
\end{array}
\label{eq20}
\end{equation}
Then, in order to decode the data matrix for any block $l$ ($k_{m-1} < l \le k_m$), we use DP to find the best estimate of $\bm{P}_j^{l}$ that maximizes $f_2 \left( \bm{P}_j^2, \cdots ,\bm{P}_j^{k_m} \right)$ in (\ref{eq20}). In order to maximize the above function, we only need to minimize $\sum_{\ell=2}^{k_m} \bm{\Lambda}_j^\ell \left( \bm{P}_j^{\ell-1},\bm{P}_j^\ell \right)$. Therefore, for any $m \geq 1$, we define the $m$th stage of the ITIC decoding using Method 2 as
\begin{equation}
\begin{array}{l}
\left \{ \bm{\hat{P}}_j^{k_{m-1}+1}, \cdots ,\bm{\hat{P}}_j^{k_m} \right \} \\
\quad \ = \displaystyle \argmin_{\bm{P}_j^{k_{m-1}+1}, \cdots ,\bm{P}_j^{k_m}} \left \{ \displaystyle \min_{\bm{P}_j^2, \cdots ,\bm{P}_j^{k_{m-1}}} \sum_{\ell=2}^{k_m} \bm{\Lambda}_j^\ell \left( \bm{P}_j^{\ell-1},\bm{P}_j^\ell \right) \right \}.
\end{array}
\label{eq21}
\end{equation}
To reduce the complexity of the exhaustive search in (\ref{eq21}), we use DP as described below. Let us denote the minimizing arguments of $\sum_{\ell=2}^{k_m} \bm{\Lambda}_j^\ell \left( \bm{P}_j^{\ell-1},\bm{P}_j^\ell \right)$ by $\bm{\hat{P}}_j^2, \cdots ,\bm{\hat{P}}_j^{k_m}$. If we know $\bm{\hat{P}}_j^{l+1}$ ($k_{m-1} < l \le k_m-1$), it can be easily shown that $\bm{\hat{P}}_j^{l}$ can be written as
\begin{equation}
\bm{\hat{P}}_j^{l} = \displaystyle \argmin_{\bm{P}_j^{l}} \left \{ \bm{\Phi}_j^{l} \left( \bm{P}_j^{l} \right) + \bm{\Lambda}_j^{l+1} \left( \bm{P}_j^{l},\bm{\hat{P}}_j^{l+1} \right) \right \}.
\label{eq23}
\end{equation}
Therefore, if we know $\bm{\hat{P}}_j^{l+1}$ and $\bm{\Phi}_j^{l} \left( \bm{P}_j^{l} \right)$, we can compute $\bm{\hat{P}}_j^{l}$ using (\ref{eq23}). This is the key element of our low complexity decoder using Method 2.

In the $m$th stage of decoding, similar to Method 1, we begin by employing (\ref{eq18}) and (\ref{eq19}) to compute and store $\bm{\Phi}_j^\ell \left( \bm{P}_j^\ell \right)$, $\ell=k_{m-1}+1, \cdots ,k_m$, for any possible data matrix $\bm{P}_j^\ell$ using the stored values of $\bm{\Phi}_j^\ell \left( \bm{P}_j^\ell \right)$ from the previous block. As in Method 1, once the signals for block $\ell$ are received, we can compute $\bm{\Phi}_j^\ell \left( \bm{P}_j^\ell \right)$ with no additional delay. Note that $\bm{\hat{P}}_j^{k_m}$ is then exactly the same as in Method 1 because (\ref{eq116}) and (\ref{eq20}) (and therefore the resulting cost functions) are identical for decoding block $l = k_m$. Thus, at block $k_m$, we compute $\bm{\hat{P}}_j^{k_m} = \argmin_{\bm{P}_j^{k_m}} \bm{\Phi}_j^{k_m} \left( \bm{P}_j^{k_m} \right)$ as the best estimate of the data matrix $\bm{P}_j^{k_m}$, which then determines the decoded bits. We then move backwards, decoding the remaining matrices one at a time beginning from $\bm{P}_j^{k_m-1}$ and ending at $\bm{P}_j^{k_{m-1}+1}$ using (\ref{eq23}), that is, utilizing the last decoded matrix and the stored values of $\bm{\Phi}_j^\ell \left( \bm{P}_j^\ell \right)$, $\ell=k_{m-1}+1, \cdots ,k_m-1$. Finally, we supply the decoded bits for each time block.


\textit{Method 3 (Decision Feedback):} An alternative approach to decoding $\bm{P}_j^l$ at block $l$ is to use the decoded matrix for $\bm{P}_j^{l-1}$ at block $l-1$ in (\ref{eq0017}). Therefore, we define the ITIC decoding using Method 3 as
\begin{equation}
\bm{\hat{P}}_j^l = \displaystyle \argmin_{\bm{P}_j^l} \bm{\Lambda}_j^l \left( \bm{\hat{P}}_j^{l-1},\bm{P}_j^l \right)
\label{eq0019}
\end{equation}
where $\bm{\hat{P}}_j^{l-1}$ is the decoded matrix for $\bm{P}_j^{l-1}$ at block $l-1$. Notice that by using this approach, in order to decode $\bm{P}_j^l$ we only need to solve an optimization over $\bm{P}_j^l$. Therefore, the decoding complexity is significantly reduced compared to the previous three decoding methods. However, the decoded signals for $\bm{P}_j^{l-1}$ at block $l-1$ may be erroneous, which can lead to error propagation and thus performance degradation. We study the effect of error propagation in Section \ref{simulations} and show that it is not significant.

\subsection{Optimal Multiple Partition Decoding Schemes}
\label{ompdecoding}

\begin{figure}
\begin{center}
  \includegraphics[scale=0.23]{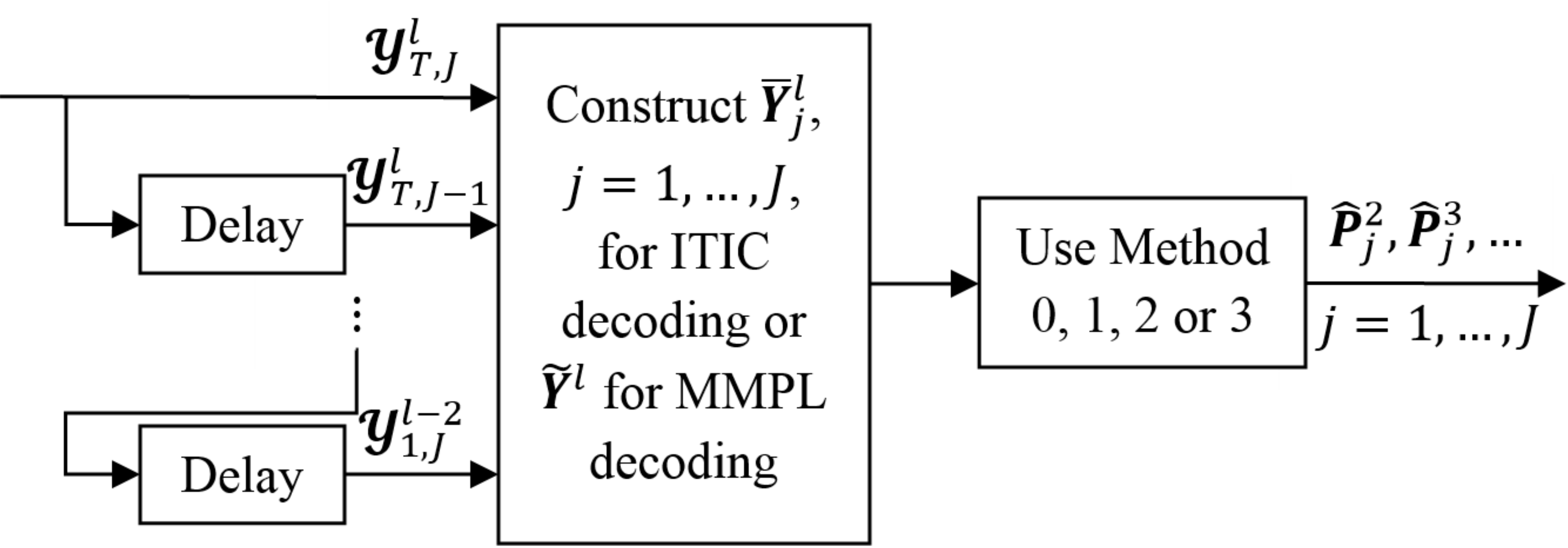}
  \caption{Block diagram of differential decoders.\label{Fig3}}
\end{center}
\end{figure}

In this subsection, we present additional decoding schemes that achieve significantly higher coding gains compared to our low complexity schemes. In order to do this, we need the following proposition:
\begin{proposition} For any $l \geq 2$, the following relationship holds
\begin{equation}
\bm{\tilde{Y}}^l = \sum_{j=1}^J \bm{H}_j \ \bm{S}_j^{l-2} \ \bm{U}_j^l \ \bm{\tilde{A}}_j + \bm{\tilde{N}}^l
\label{eq0020}
\end{equation}
where $\bm{\tilde{A}}_j$ is a $3T \times 3TJ-J+1$ matrix given by
\setcounter{MaxMatrixCols}{11}
\begin{equation}
\small{\begin{array} {c}
\bm{\tilde{A}}_j = \begin{pmatrix}[1.9]
               \smash{\overbrace{1 \cdots 1}^{j\ \textrm{times}}} & 0 \cdots 0 & 0 \cdots 0 & \cdots & 0 \cdots 0 \\
               0 \cdots 0 & \smash{\overbrace{1 \cdots 1}^{J\ \textrm{times}}} & 0 \cdots 0 & \cdots & 0 \cdots 0 \\
               0 \cdots 0 & 0 \cdots 0 & \smash{\overbrace{1 \cdots 1}^{J\ \textrm{times}}} & \cdots & 0 \cdots 0 \\
               \ddots & \ddots & \ddots & \ddots & \ddots \\
               0 \cdots 0 & 0 \cdots 0 & 0 \cdots 0 & \cdots & \smash{\overbrace{1 \cdots 1}^{J-j+1 \textrm{times}}}
           \end{pmatrix}, \\\\
\begin{array} {l}
\bm{\tilde{Y}}^l = \left(\bm{y}_{1,J}^{l-2}, \bm{y}_{2,1}^{l-2}, \bm{y}_{2,2}^{l-2}, \cdots, \bm{y}_{T,J}^{l-2}, \bm{y}_{1,1}^{l-1}, \right. \\
\qquad \qquad \qquad \qquad \qquad \ \left.\bm{y}_{1,2}^{l-1}, \cdots, \bm{y}_{T,J}^{l-1}, \bm{y}_{1,1}^l, \bm{y}_{1,2}^l, \cdots, \bm{y}_{T,J}^l\right), \\
\bm{\tilde{N}}^l = \left(\bm{n}_{1,J}^{l-2}, \bm{n}_{2,1}^{l-2}, \bm{n}_{2,2}^{l-2}, \cdots, \bm{n}_{T,J}^{l-2}, \bm{n}_{1,1}^{l-1}, \right. \\
\qquad \qquad \qquad \qquad \qquad \left.\bm{n}_{1,2}^{l-1}, \cdots, \bm{n}_{T,J}^{l-1}, \bm{n}_{1,1}^l, \bm{n}_{1,2}^l, \cdots, \bm{n}_{T,J}^l\right),
\end{array} \\
\bm{U}_j^l = \left(\bm{I}_N, \bm{P}_j^{l-1}, \bm{P}_j^{l-1}\bm{P}_j^l\right), \ j=1,\cdots,J.
\end{array}}
\label{eq0021}
\end{equation}
\begin{proof}
The result follows from the input-output relationship for any time block $l>0$ available in Appendix A and using simple algebra.
\end{proof}
\end{proposition}
Again, notice that $\bm{\tilde{Y}}^l$ starts from $\bm{y}_{1,J}^{l-2}$ instead of $\bm{y}_{1,1}^{l-2}$. Other previously received signals could be considered to improve performance, but that would cause additional inter-block interference from previously transmitted signals and would increase decoding complexity. It is easy to see from Proposition 4.2 that when conditioned on the data matrices $\bm{P}_1^{l-1}, \bm{P}_1^l, \cdots, \bm{P}_J^{l-1}, \bm{P}_J^l$, the matrix $\bm{\tilde{Y}}^l$ is Gaussian with conditional pdf
\begin{equation}
\footnotesize{P\left(\bm{\tilde{Y}}^l \left| \bm{P}_1^{l-1}, \bm{P}_1^l, \cdots, \bm{P}_J^{l-1}, \bm{P}_J^l\right.\right) \propto \frac{\exp\left\{-\Tr\left[\bm{\tilde{Y}}^l\cdot(\bm{\tilde{V}}^l)^{-1}\cdot(\bm{\tilde{Y}}^l)^\dagger\right]\right\}}{\left[\det(\bm{\tilde{V}}^l)\right]^M}}
\label{eq0022}
\end{equation}
where $\bm{\tilde{V}}^l$ is the covariance matrix given by $\bm{\tilde{V}}^l = \sum_{j=1}^J (\bm{U}_j^l \ \bm{\tilde{A}}_j)^\dagger \cdot (\bm{U}_j^l \ \bm{\tilde{A}}_j) + (\textrm{SNR})^{-1} T_s \cdot \bm{\tilde{D}}$ and $\bm{\tilde{D}} = \diag(\tau_{1,0}^{-1}, \tau_{2,1}^{-1}, \cdots, \tau_{3TJ-J+1,3TJ-J}^{-1})$ is a $3TJ-J+1 \times 3TJ-J+1$ diagonal matrix. Based on (\ref{eq0022}), we can define the Maximum Multiple Partition Likelihood (MMPL) decoding using Method 0 as
\begin{equation}
\footnotesize{\begin{array} {l}
\left \{ \bm{\hat{P}}_1^{l-1}, \bm{\hat{P}}_1^l, \cdots, \bm{\hat{P}}_J^{l-1}, \bm{\hat{P}}_J^l \right \} \\
= \displaystyle \argmin_{\bm{P}_1^{l-1}, \bm{P}_1^l, \cdots, \bm{P}_J^{l-1}, \bm{P}_J^l} \left \{ M\cdot\ln\left[\det(\bm{\tilde{V}}^l)\right] + \Tr\left[\bm{\tilde{Y}}^l\cdot(\bm{\tilde{V}}^l)^{-1}\cdot(\bm{\tilde{Y}}^l)^\dagger\right] \right \}.
\end{array}}
\label{eq0023}
\end{equation}
The cost function of the MMPL decoder using Method 0 is a function of $\bm{P}_1^{l-1}, \bm{P}_1^l, \cdots, \bm{P}_J^{l-1}, \bm{P}_J^l$, whereas the cost function of the ITIC decoder for User $j=1,\cdots,J$ using Method 0 is only a function of $\bm{P}_j^{l-1}, \bm{P}_j^l$. We can use the DP procedures in Methods 1 and 2 with the cost function of the MMPL decoder in (\ref{eq0023}) just as with the cost function of the ITIC decoder in (\ref{eq0017}). However, we need to compute and store a function of $\bm{P}_1^l, \cdots, \bm{P}_J^l$ instead of $\bm{\Phi}_j^l \left( \bm{P}_j^l \right)$ defined in (\ref{eq18}). Similarly, Method 3 can be applied to the cost function of the MMPL decoder in (\ref{eq0023}) by using the decoded matrices for $\bm{P}_1^{l-1}, \cdots, \bm{P}_J^{l-1}$ at block $l-1$ in (\ref{eq0023}) to decode $\bm{P}_1^l, \cdots, \bm{P}_J^l$ at block $l$. The three algorithms can therefore be changed accordingly. The block diagram of the proposed differential decoders is shown in Fig. \ref{Fig3}.

The corresponding coherent decoders for the ITIC and MMPL decoders can be derived using similar procedures to the ones described above as well.
Due to space limitations, we do not provide the details of the coherent ITIC and MMPL decoders.

\section{Diversity Analysis}
\label{diversitygainanalysis}

With a small abuse of the notation, for data matrices $\bm{P}_1, \bm{P}_2, \bm{P}_3, \bm{P}_4$, let us define
\begin{equation}
\bm{G}(\bm{P}_1, \bm{P}_2, \bm{P}_3, \bm{P}_4) \triangleq \begin{pmatrix}[0.75]
               \bm{I}_N & \bm{P}_1 & \bm{P}_1 \bm{P}_2 \\
               \bm{I}_N & \bm{P}_3 & \bm{P}_3 \bm{P}_4
           \end{pmatrix} \cdot \bm{\bar{A}}
\label{eq0314}
\end{equation}
where $\bm{\bar{A}}$ is the $3T \times 3T-1$ matrix given in (\ref{eq0015}).
Suppose that we choose the signal constellation such that for any possible data matrices $\bm{P}_1, \bm{P}_2, \bm{P}_3, \bm{P}_4$ with $(\bm{P}_1, \bm{P}_2) \ne (\bm{P}_3, \bm{P}_4)$, the matrix $\bm{G}(\bm{P}_1, \bm{P}_2, \bm{P}_3, \bm{P}_4)$ has full row rank (i.e., $\bm{G}(\bm{P}_1, \bm{P}_2, \bm{P}_3, \bm{P}_4)$ is of rank $2N$). We prove that under this condition all the proposed schemes achieve a diversity order of $MN$ (full diversity). We also derive an equivalent condition, which can be easily verified using simple matrix operations. Furthermore, for the cases of two and four transmit antennas, we provide examples of PSK constellations to achieve full diversity.
\begin{theorem}
The proposed ITIC and MMPL decoders using Method 0 achieve full diversity.
\begin{proof}
See Appendix B.
\end{proof}
\end{theorem}

The following theorem extends the result of Theorem 5.1 to all the proposed methods:
\begin{theorem}
If one of the proposed differential schemes using Method 0 provides full diversity, then the corresponding differential schemes using Methods 1, 2 and 3 will provide full diversity as well.
\begin{proof}
The proof is very similar to that of Theorem 5.1 in \cite{PJ13}.
\end{proof}
\end{theorem}
Therefore, by Theorems 5.1 and 5.2, all the proposed differential schemes (i.e., ITIC and MMPL decoders using Methods 0, 1, 2 and 3) provide full diversity.

As mentioned above, in order to guarantee full diversity, we need to make sure that $\bm{G}(\bm{P}_1, \bm{P}_2, \bm{P}_3, \bm{P}_4)$ has full row rank for any possible data matrices $\bm{P}_1, \bm{P}_2, \bm{P}_3, \bm{P}_4$ with $(\bm{P}_1, \bm{P}_2) \ne (\bm{P}_3, \bm{P}_4)$. In the following theorem we derive an equivalent condition, which can be easily verified using simple matrix operations:

\begin{theorem}
$\bm{G}(\bm{P}_1, \bm{P}_2, \bm{P}_3, \bm{P}_4)$ has full row rank for any possible data matrices $\bm{P}_1, \bm{P}_2, \bm{P}_3, \bm{P}_4$ with $(\bm{P}_1, \bm{P}_2) \ne (\bm{P}_3, \bm{P}_4)$ if and only if
\begin{equation}
\scriptsize{\bm{w} \cdot \left[\bm{\tilde{P}}_1\bm{\tilde{P}}_2 + N \left(\bm{I}_N - \bm{\tilde{P}}_1\right) \cdot \left(\frac{\left(\bm{\tilde{P}}_3 - \bm{\tilde{P}}_1\right)^\dagger}{\big\| \bm{\tilde{P}}_3 - \bm{\tilde{P}}_1 \big\|_F^2}\right) \cdot \left(\bm{\tilde{P}}_3\bm{\tilde{P}}_4 - \bm{\tilde{P}}_1\bm{\tilde{P}}_2\right)\right] \ne \bm{w}}
\label{eq0340}
\end{equation}
for any possible data matrices $\bm{\tilde{P}}_1, \bm{\tilde{P}}_2, \bm{\tilde{P}}_3, \bm{\tilde{P}}_4$ with $\bm{\tilde{P}}_1 \ne \bm{\tilde{P}}_3$, \\\\
where $\bm{w} = (\smash{\overbrace{1,1,\cdots,1}^{N \textrm{times}}})$.
\begin{proof}
See Appendix C.
\end{proof}
\end{theorem}
For instance, consider the case when the Alamouti code is used to construct the data matrices $\bm{P}_j^l$. Then, one can use Theorem 5.3 to verify that when the BPSK constellation $\left\{\frac{e^{j(\frac{\pi}{4})}}{\sqrt{2}}, -\frac{e^{j(\frac{\pi}{4})}}{\sqrt{2}}\right\}$ or the QPSK constellation $\left\{\frac{e^{j(\frac{\pi}{8})}}{\sqrt{2}}, j\left(\frac{e^{j(\frac{\pi}{8})}}{\sqrt{2}}\right), -\frac{e^{j(\frac{\pi}{8})}}{\sqrt{2}}, -j\left(\frac{e^{j(\frac{\pi}{8})}}{\sqrt{2}}\right)\right\}$ is used, $\bm{G}(\bm{P}_1, \bm{P}_2, \bm{P}_3, \bm{P}_4)$ will have full row rank for any possible data matrices $\bm{P}_1, \bm{P}_2, \bm{P}_3, \bm{P}_4$ with $(\bm{P}_1, \bm{P}_2) \ne (\bm{P}_3, \bm{P}_4)$. As another example, consider the case when the following $4 \times 4$ rate-one STBC \cite{J05} is used to construct the data matrices:
\begin{equation}
\bm{P}_j^l = \begin{pmatrix}[0.75]
               p_{j,1}^l & -p_{j,2}^l & -p_{j,3}^l & -p_{j,4}^l \\
               p_{j,2}^l & p_{j,1}^l & p_{j,4}^l & -p_{j,3}^l \\
               p_{j,3}^l & -p_{j,4}^l & p_{j,1}^l & p_{j,2}^l \\
               p_{j,4}^l & p_{j,3}^l & -p_{j,2}^l & p_{j,1}^l
           \end{pmatrix}.
\label{eq0341}
\end{equation}
Note that the above STBC is orthogonal for the BPSK constellation $\left\{\frac{e^{j(\frac{\pi}{4})}}{2}, -\frac{e^{j(\frac{\pi}{4})}}{2}\right\}$. Again, one may use Theorem 5.3 to verify that when the BPSK constellation $\left\{\frac{e^{j(\frac{\pi}{4})}}{2}, -\frac{e^{j(\frac{\pi}{4})}}{2}\right\}$ is used, $\bm{G}(\bm{P}_1, \bm{P}_2, \bm{P}_3, \bm{P}_4)$ will have full row rank for any possible data matrices $\bm{P}_1, \bm{P}_2, \bm{P}_3, \bm{P}_4$ with $(\bm{P}_1, \bm{P}_2) \ne (\bm{P}_3, \bm{P}_4)$.

\section{Simulation Results}
\label{simulations}

\begin{figure}
\begin{center}
  \includegraphics[scale=0.65]{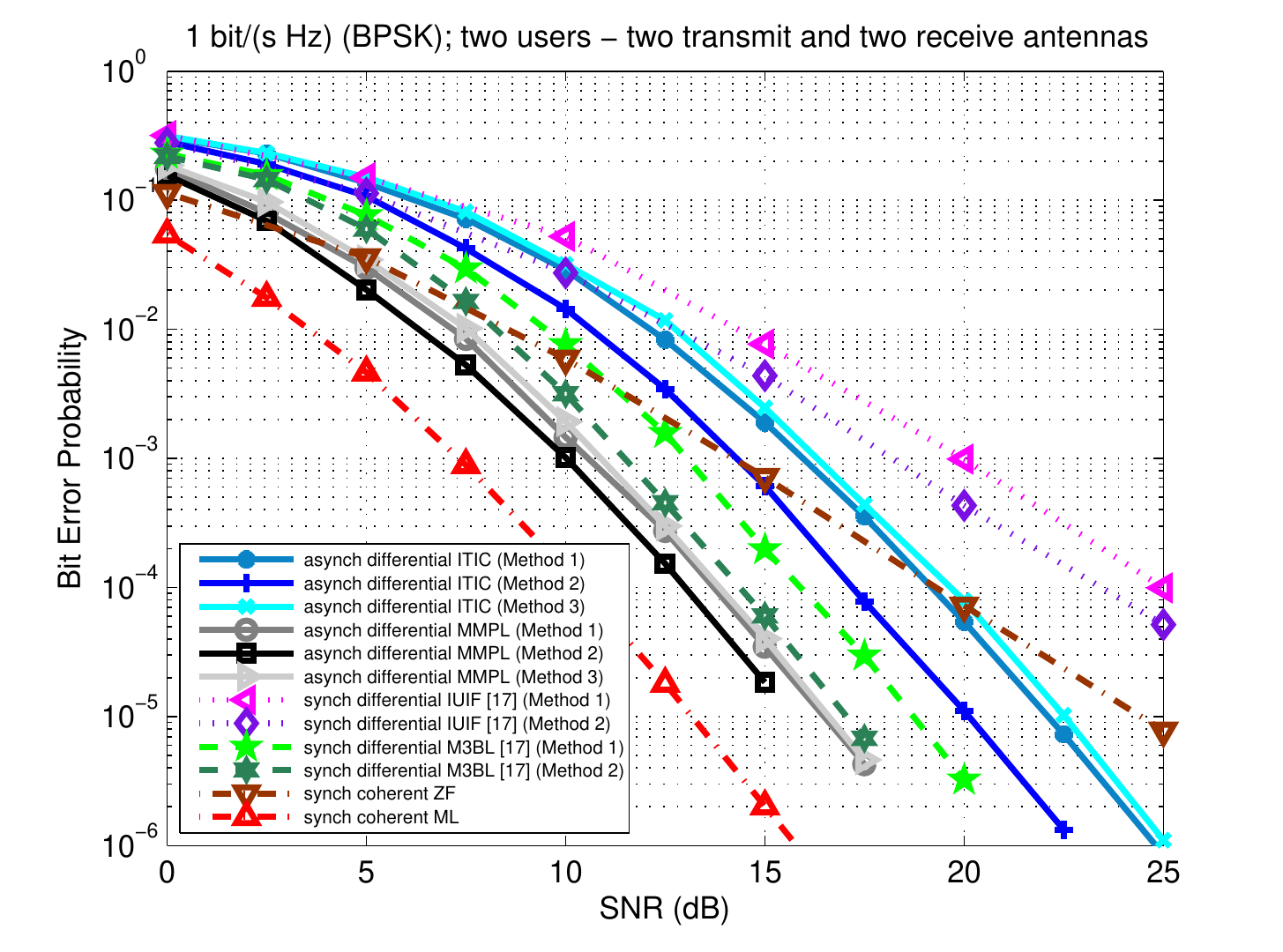}
  \caption{Performance of the proposed asynchronous differential schemes for $\tau_2-\tau_1=T_s/2$, the synchronous differential schemes in \cite{PJ13}, and the synchronous coherent schemes using ZF and ML decoding at a rate of 1 b/(s Hz) for 2 users each with 2 transmit antennas and 1 receiver with 2 receive antennas.\label{Fig4}}
\end{center}
\end{figure}

\begin{figure}
\begin{center}
  \includegraphics[scale=0.65]{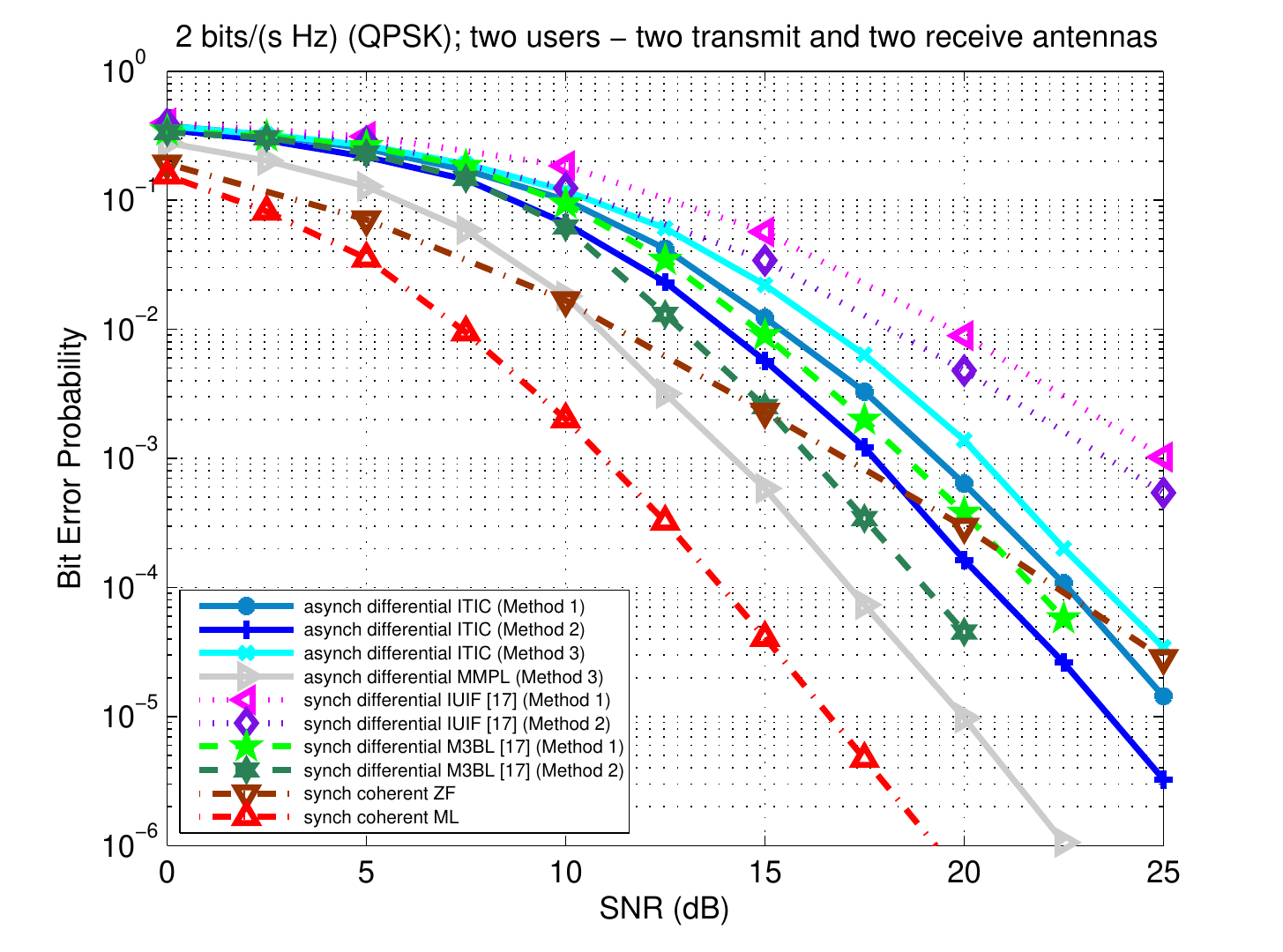}
  \caption{Performance of the proposed asynchronous differential schemes for $\tau_2-\tau_1=T_s/2$, the synchronous differential schemes in \cite{PJ13}, and the synchronous coherent schemes using ZF and ML decoding at a rate of 2 b/(s Hz) for 2 users each with 2 transmit antennas and 1 receiver with 2 receive antennas.\label{Fig5}}
\end{center}
\end{figure}

\begin{figure}
\begin{center}
  \includegraphics[scale=0.65]{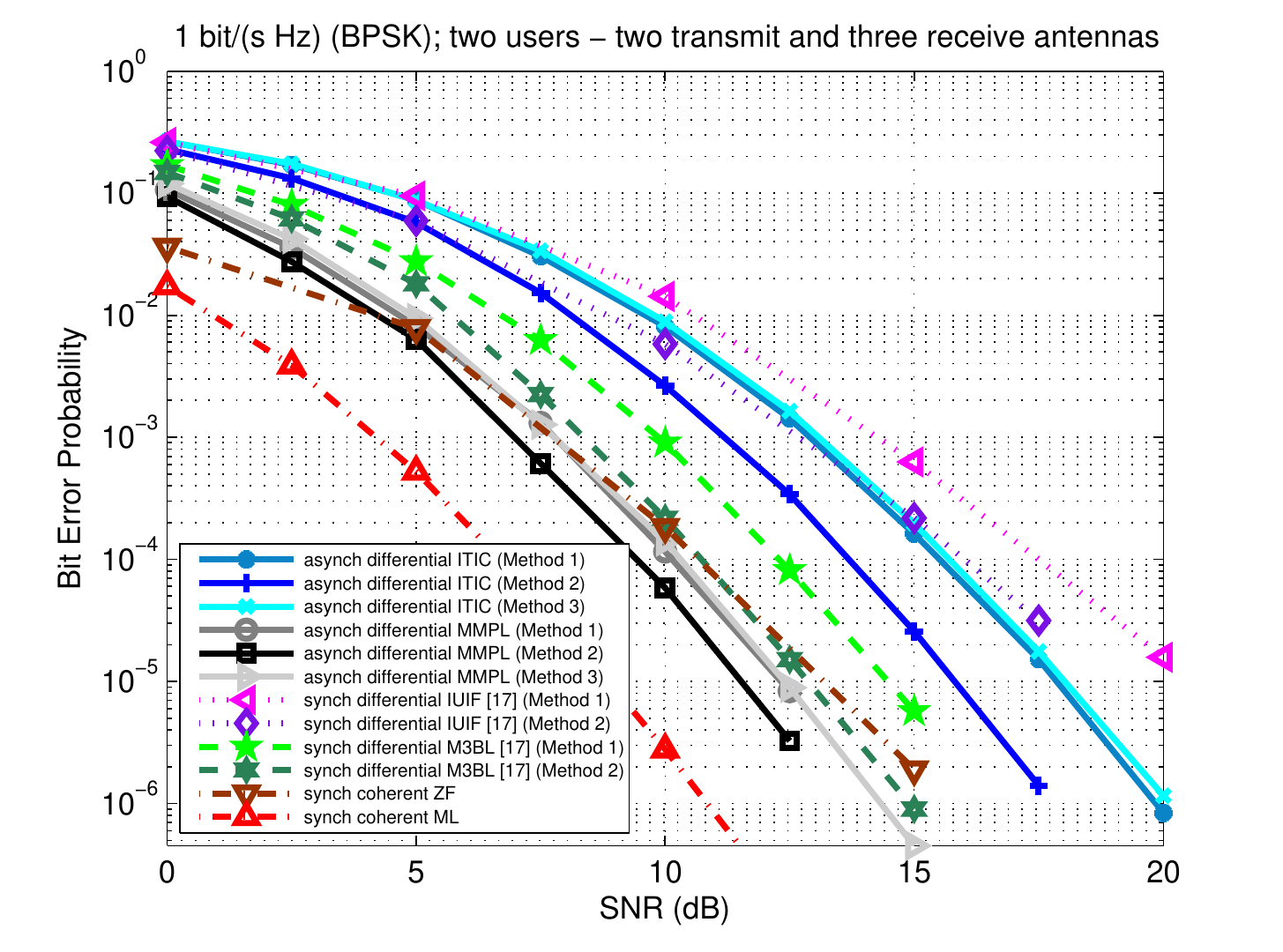}
  \caption{Performance of the proposed asynchronous differential schemes for $\tau_2-\tau_1=T_s/2$, the synchronous differential schemes in \cite{PJ13}, and the synchronous coherent schemes using ZF and ML decoding at a rate of 1 b/(s Hz) for 2 users each with 2 transmit antennas and 1 receiver with 3 receive antennas.\label{Fig6}}
\end{center}
\end{figure}

\begin{figure}
\begin{center}
  \includegraphics[scale=0.65]{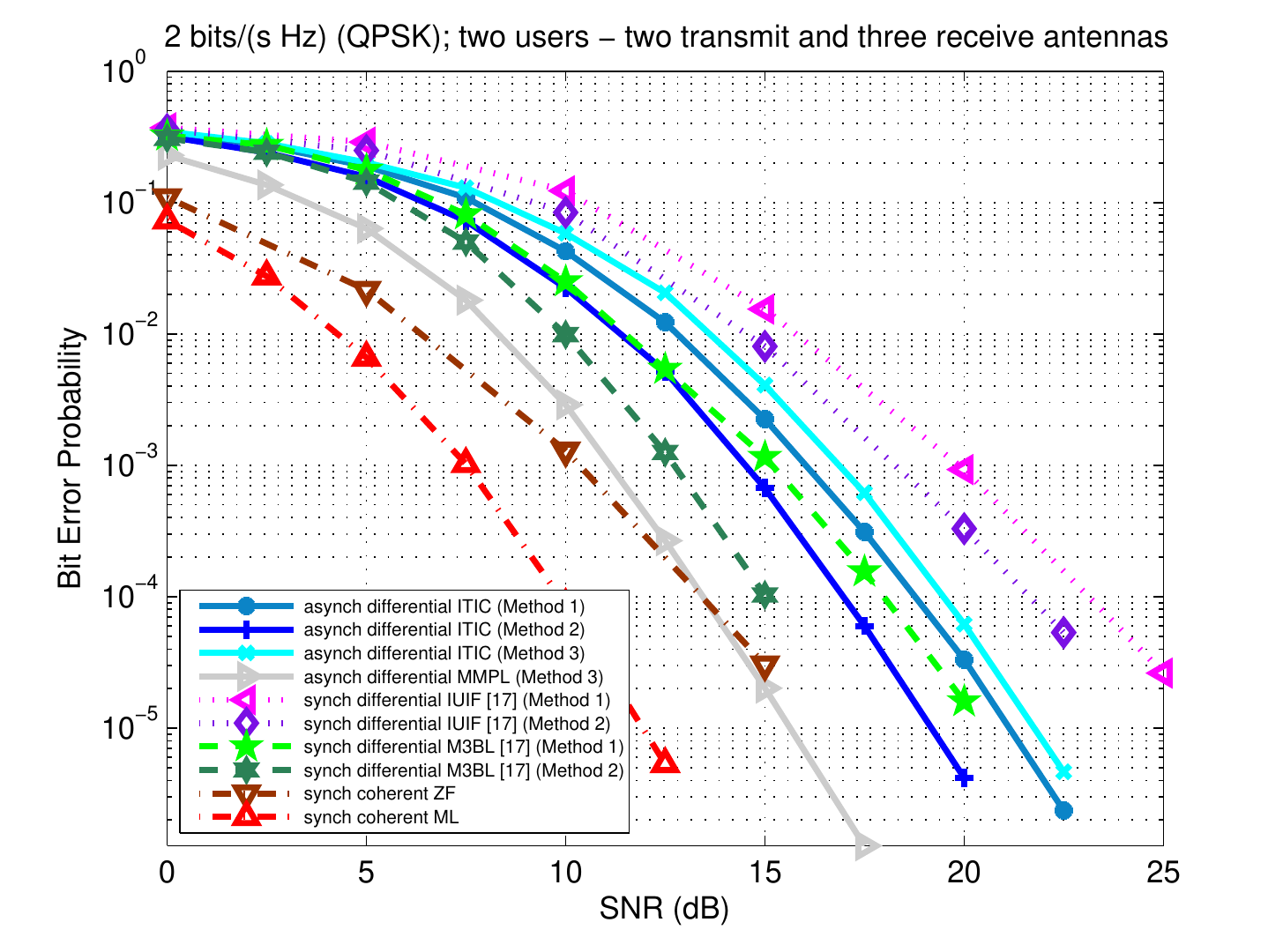}
  \caption{Performance of the proposed asynchronous differential schemes for $\tau_2-\tau_1=T_s/2$, the synchronous differential schemes in \cite{PJ13}, and the synchronous coherent schemes using ZF and ML decoding at a rate of 2 b/(s Hz) for 2 users each with 2 transmit antennas and 1 receiver with 3 receive antennas.\label{Fig7}}
\end{center}
\end{figure}

\begin{figure}
\begin{center}
  \includegraphics[scale=0.65]{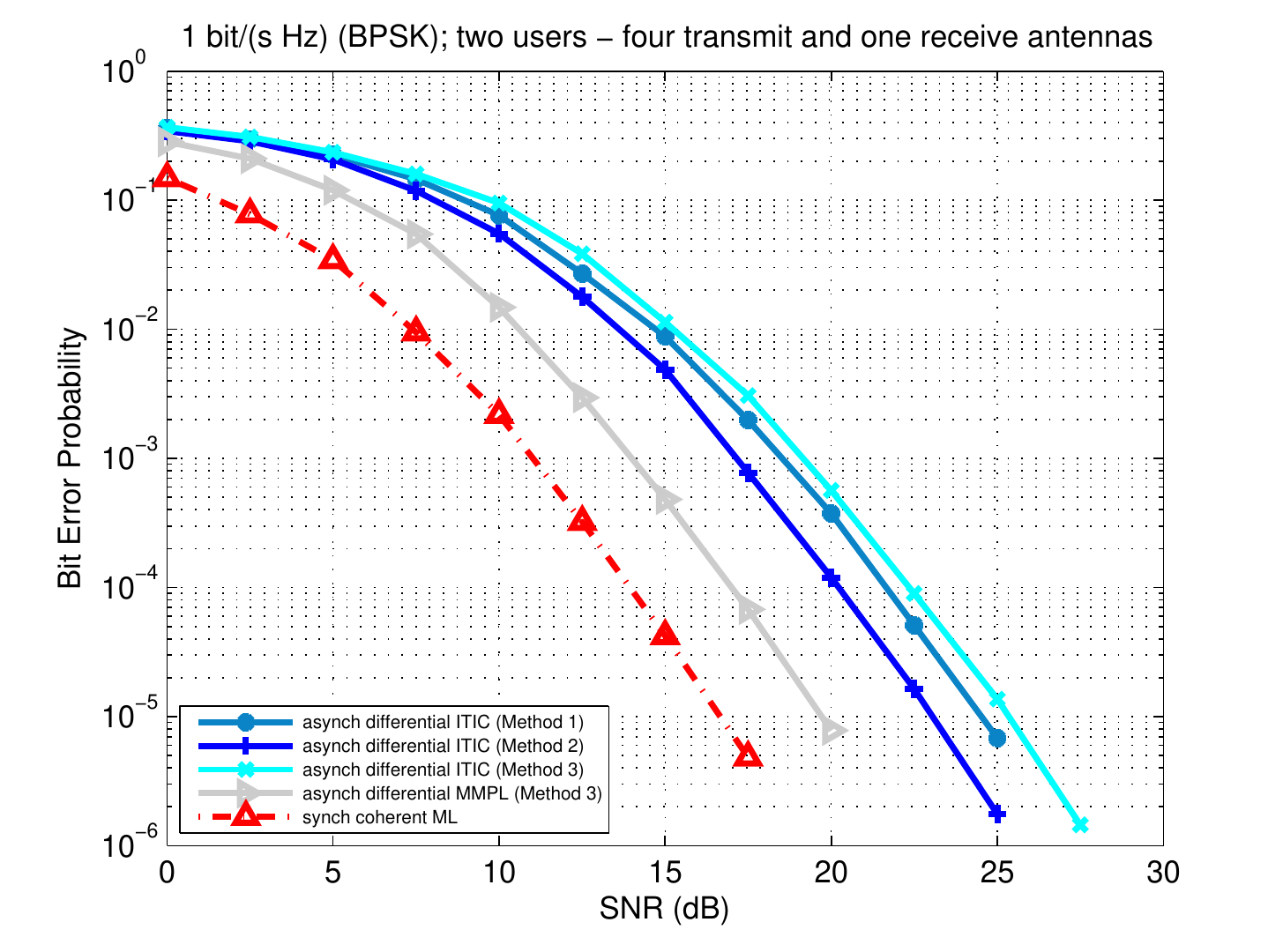}
  \caption{Performance of the proposed asynchronous differential schemes for $\tau_2-\tau_1=T_s/2$ and the synchronous coherent scheme using ML decoding at a rate of 1 b/(s Hz) for 2 users each with 4 transmit antennas and 1 receiver with 1 receive antenna.\label{Fig8}}
\end{center}
\end{figure}

\begin{figure}
\begin{center}
  \includegraphics[scale=0.65]{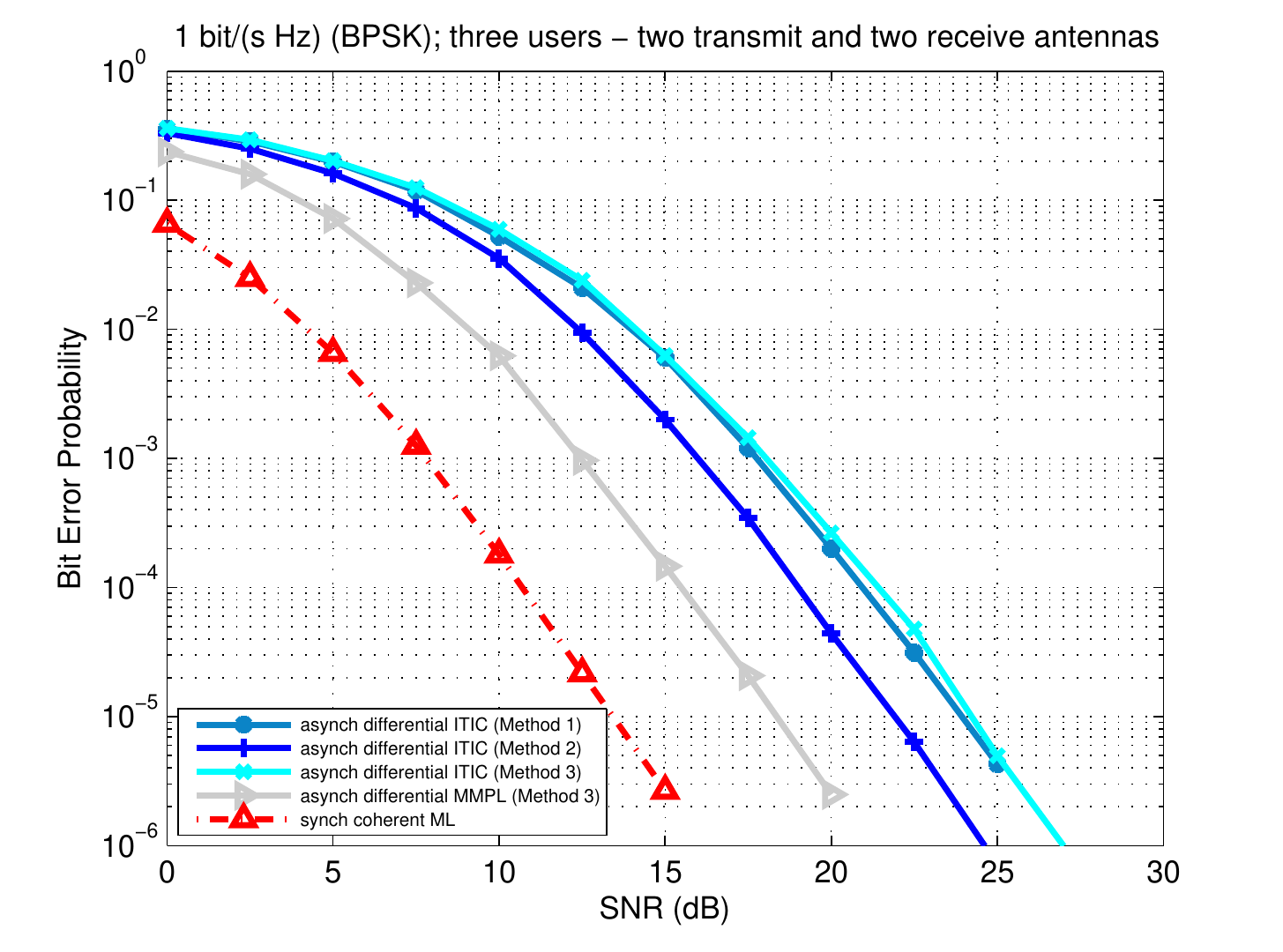}
  \caption{Performance of the proposed asynchronous differential schemes for $\tau_{j+1}-\tau_j=T_s/3$, $\forall j$, and the synchronous coherent scheme using ML decoding at a rate of 1 b/(s Hz) for 3 users each with 2 transmit antennas and 1 receiver with 2 receive antennas.\label{Fig9}}
\end{center}
\end{figure}

\begin{figure}
\begin{center}
  \includegraphics[scale=0.65]{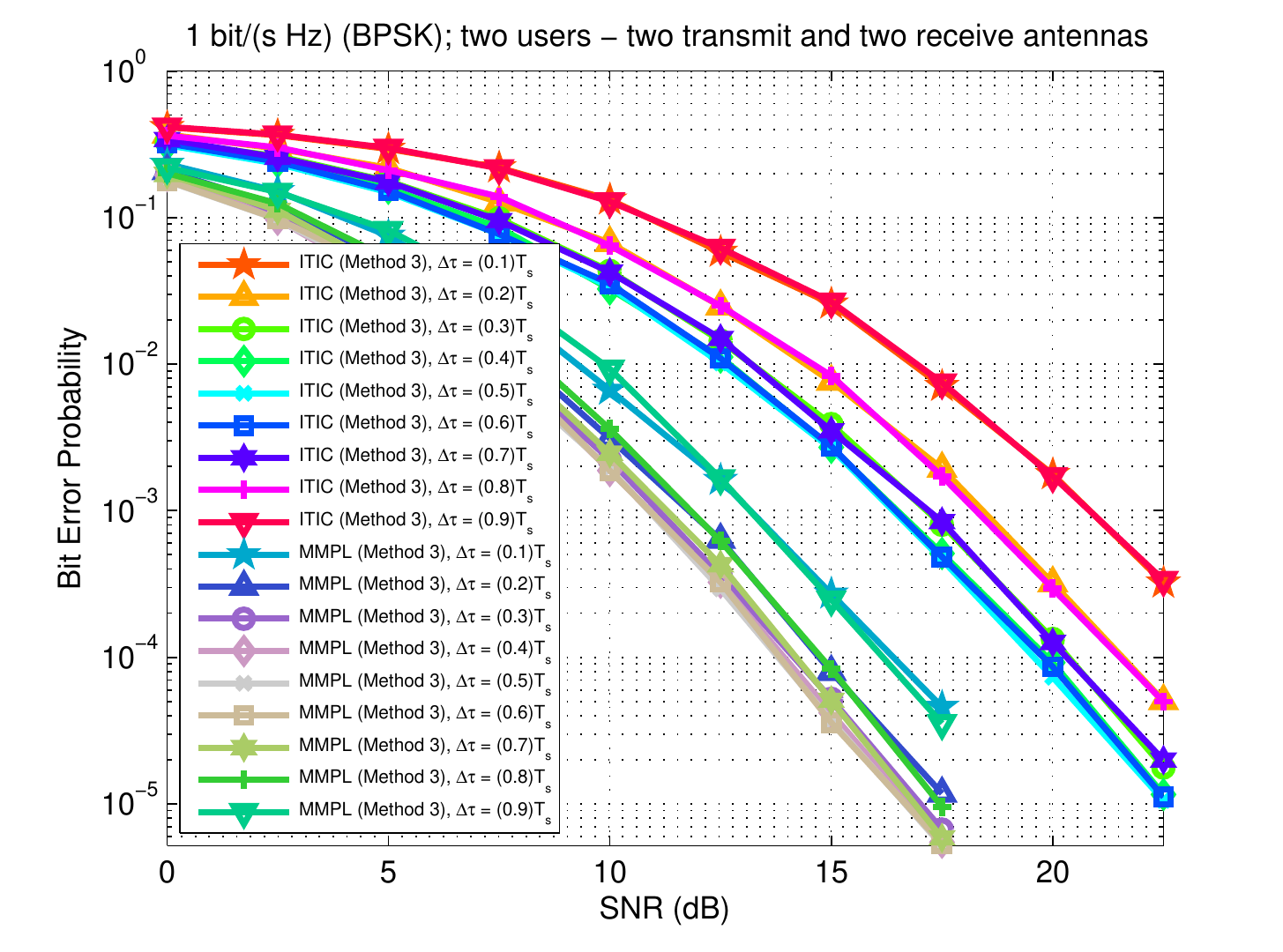}
  \caption{Comparison of the proposed asynchronous differential schemes using Method 3 for different relative time delays $\Delta\tau=\tau_2-\tau_1$ at a rate of 1 \qquad b/(s Hz) for 2 users each with 2 transmit antennas and 1 receiver with 2 receive antennas.\label{Fig10}}
\end{center}
\end{figure}

In this section, we provide simulation results for the performance of the proposed differential modulation schemes using the ITIC and MMPL decoders based on Methods 1, 2 and 3. We compare the performance of our schemes to the IUIF and M3BL differential schemes presented in \cite{PJ13} and the synchronous coherent schemes using Zero-Forcing (ZF) and ML decoding. When using Method 2 for decoding, we decode all the signals within each frame after receiving the last signal in that frame. In our simulations, the channel is quasi-static flat Rayleigh fading where the fading is constant within one frame and varies independently from one frame to another. Depending on the number of transmit antennas, we use either the Alamouti code or the $4 \times 4$ OSTBC in (\ref{eq0341}) for all users to encode and transmit 64 data matrices per user in each frame.
Also, we use the BPSK and QPSK constellations described in Section \ref{diversitygainanalysis} as the signal constellations for the simulations of our differential schemes at transmission rates 1 b/(s Hz) and 2 b/(s Hz), respectively. In Figs. \ref{Fig4}-\ref{Fig9}, we consider the relative time delays between the received signals of consecutive users to be equal (i.e., $\tau_{j+1}-\tau_j=T_s/J$, $\forall j$). We study the effect of other relative time delays on performance in Fig. \ref{Fig10}. In each figure, the curves for all users are identical.

Figs. \ref{Fig4} and \ref{Fig5} show BER as a function of SNR at transmission rates 1 b/(s Hz) and 2 b/(s Hz), respectively, for 2 users each equipped with 2 transmit antennas and a receiver with 2 receive antennas.
In Figs. \ref{Fig6} and \ref{Fig7}, we present similar results for 3 receive antennas.
In Fig. \ref{Fig8}, we provide simulation results at a transmission rate of 1 b/(s Hz) for 2 users each equipped with 4 transmit antennas and a receiver with 1 receive antenna.
Note that all our schemes work for any number of receive antennas, while the low complexity differential schemes in \cite{PJ13} require at least $J$ receive antennas. All simulation results demonstrate that all the proposed schemes achieve full diversity like the corresponding coherent schemes using ML decoding. On the other hand, the low complexity differential schemes in \cite{PJ13} only provide full transmit diversity. Additionally, compared to the differential schemes in \cite{PJ13}, the MMPL decoding schemes provide significant performance improvement. Therefore, the proposed schemes provide the possibility of a tradeoff between decoding complexity and the coding gain.

In Fig. \ref{Fig9}, we show BER as a function of SNR at a transmission rate of 1 b/(s Hz) for 3 users each equipped with 2 transmit antennas and a receiver with 2 receive antennas.
With the assumption of equal relative time delays, it can be seen from Proposition 4.1 and the covariance matrices for the noise vectors given in Section \ref{systemmodel} that the effect of changing the number of users from $J_1$ to $J_2$ on the performance of the ITIC decoders is the same as that of multiplying the SNR by $J_1/J_2$. This corresponds to a change of $10 \log_{10}(J_1/J_2)$ dB in performance. As expected, the performances of the ITIC decoders in Fig. \ref{Fig4} for 2 users are $10 \log_{10}(3/2) \approx 1.8$ dB better than those of Fig. \ref{Fig9} for 3 users.
All simulations show that the effect of error propagation on the performance of the proposed schemes using Method 3 is very small. Our schemes using Method 3 have lower decoding complexity compared to their corresponding schemes using Method 1, yet the proposed schemes using Method 3 provide almost the same performance as their corresponding schemes using Method 1.

Finally, we compare the performance of our differential schemes with different relative time delays between the received signals. Again, we consider a system with 2 users each equipped with 2 transmit antennas and a receiver with 2 receive antennas.
Fig. \ref{Fig10} shows the performance of the ITIC and MMPL decoders using Method 3 for different values of $\Delta\tau=\tau_2-\tau_1$ at a transmission rate of 1 b/(s Hz).
The results for our decoding schemes using Methods 0, 1 and 2 are similar. It is evident from the simulations that the proposed schemes perform best when $\Delta\tau=T_s/2$, that is, when the signals of the two users are received with a time difference of half a symbol. Moreover, for values of $\Delta\tau$ close to $T_s/2$, the performance of our schemes is close to the best performance for $\Delta\tau=T_s/2$ and deviates from the best performance more quickly as $\Delta\tau$ deviates from $T_s/2$. This is in line with capacity results reported in \cite{V89} where $\Delta\tau=T_s/2$ provides the highest value of channel capacity in a two-user MAC.

\section{Conclusion}
\label{conclusions}

We introduced differential detection schemes for asynchronous multi-user MIMO systems based on orthogonal STBCs where neither the transmitters nor the receiver knows the CSI. We first presented schemes with simple differential encoding and low complexity differential decoding algorithms by performing interference cancelation in time and employing different decoding methods. The decoding complexity of these schemes increases linearly with the number of users. We then presented additional differential decoding schemes that achieve significantly higher coding gains compared to our low complexity schemes. Simulation results show that they also outperform the existing synchronous differential schemes. The proposed schemes work for any square OSTBC, any constant amplitude constellation, any number of users, and a receiver with any number of receive antennas. Similar to the case of a single user, our schemes can be extended to work with other STBCs with higher rates, such as QOSTBCs, through minor changes. Furthermore, we derived conditions under which our schemes provide full diversity. For the cases of two and four transmit antennas, we also provided examples of PSK constellations to achieve full diversity. To the best of our knowledge, the proposed differential modulation schemes are the first differential schemes for asynchronous multi-user communication systems.


%

\appendices
\section{Proof of Proposition 4.1}
Using the input-output relationship in (\ref{eq0008}) and (\ref{eq0012}), we can write the input-output relationship for a single time block $l>0$ as
\begin{equation}
\footnotesize{\begin{array}{l}
\begin{pmatrix}
    \bm{y}_{1,1}^l, \cdots, \bm{y}_{1,J}^l, \cdots, \bm{y}_{T,1}^l, \cdots, \bm{y}_{T,J}^l
\end{pmatrix} \\
= \displaystyle\sum_{i=1}^J \bm{H}_i \begin{pmatrix}
    \bm{S}_i^{l-1}, \bm{S}_i^l
\end{pmatrix}\begin{pmatrix}
    \bm{Z}_{i,1} \\
    \bm{Z}_{i,0}
\end{pmatrix} + \begin{pmatrix}
    \bm{n}_{1,1}^l, \cdots, \bm{n}_{1,J}^l, \cdots, \bm{n}_{T,1}^l, \cdots, \bm{n}_{T,J}^l
\end{pmatrix}
\end{array}}
\label{eq0214}
\end{equation}
where $\bm{Z}_{i,0},\bm{Z}_{i,1}$, $i=1,\cdots,J$, are $T \times TJ$ matrices given by
\begin{equation}
\footnotesize{\begin{array} {c}
\bm{Z}_{i,0} = \begin{pmatrix}[1.9]
               \smash{\overbrace{0 \cdots 0}^{i-1 \textrm{times}}} & \smash{\overbrace{1 \cdots 1}^{J\ \textrm{times}}} & 0 \cdots 0 & \cdots & 0 \cdots 0 \\
               0 \cdots 0 & 0 \cdots 0 & \smash{\overbrace{1 \cdots 1}^{J\ \textrm{times}}} & \cdots & 0 \cdots 0 \\
               \ddots & \ddots & \ddots & \ddots & \ddots \\
               0 \cdots 0 & 0 \cdots 0 & 0 \cdots 0 & \cdots & \smash{\overbrace{1 \cdots 1}^{J-i+1 \textrm{times}}}
           \end{pmatrix}, \\
\bm{Z}_{i,1} = \begin{pmatrix}[1.9]
               \ 0 \cdots 0 & 0 \cdots 0 & 0 \cdots 0 & \ \ \cdots \ \ & 0 \cdots 0 \ \\
               \ 0 \cdots 0 & 0 \cdots 0 & 0 \cdots 0 & \ \ \cdots \ \ & 0 \cdots 0 \ \\
               \ \ddots & \ddots & \ddots & \ddots & \ddots \ \\
               \undermat{i-1\ \textrm{times}}{\ 1 \cdots 1} & \undermat{TJ-i+1\ \textrm{times}}{0 \cdots 0 & 0 \cdots 0 & \ \ \cdots \ \ & 0 \cdots 0} \
           \end{pmatrix}.\\\\
\end{array}}
\label{eq0215}
\end{equation}
Then, note that the interference of all users on User $j$ can be canceled by subtracting $\bm{y}_{t,j-1}^l$ from $\bm{y}_{t,j}^l$ for $t=1,\cdots,T$ as follows
\begin{equation}
\small{\begin{array} {l}
\begin{pmatrix}\bm{y}_{1,j}^l-\bm{y}_{1,j-1}^l, \cdots, \bm{y}_{T,j}^l-\bm{y}_{T,j-1}^l\end{pmatrix} \\
= \bm{H}_j \begin{pmatrix}
    \bm{S}_j^{l-1}, \bm{S}_j^l
\end{pmatrix}\begin{pmatrix}
    \bm{\bar{Z}}_1 \\
    \bm{\bar{Z}}_0
\end{pmatrix}
+ \begin{pmatrix}\bm{n}_{1,j}^l-\bm{n}_{1,j-1}^l, \cdots, \bm{n}_{T,j}^l-\bm{n}_{T,j-1}^l\end{pmatrix}
\end{array}}
\label{eq0216}
\end{equation}
where $\bm{\bar{Z}}_0,\bm{\bar{Z}}_1$ are $T \times T$ matrices given by
\begin{equation}
\footnotesize{\begin{array} {c}
\bm{\bar{Z}}_0 = \begin{pmatrix}[0.75]
               1 & -1 & 0 & \cdots & 0 & 0 \\
               0 & 1 & -1 & \cdots & 0 & 0 \\
               0 & 0 & 1 & \cdots & 0 & 0 \\
               \ddots & \ddots & \ddots & \ddots & \ddots & \ddots \\
               0 & 0 & 0 & \cdots & -1 & 0 \\
               0 & 0 & 0 & \cdots & 1 & -1 \\
               0 & 0 & 0 & \cdots & 0 & 1
           \end{pmatrix}, \\\\
\bm{\bar{Z}}_1 = \begin{pmatrix}[0.75]
               0 & 0 & 0 & \cdots & 0 & 0 \\
               0 & 0 & 0 & \cdots & 0 & 0 \\
               0 & 0 & 0 & \cdots & 0 & 0 \\
               \ddots & \ddots & \ddots & \ddots & \ddots & \ddots \\
               0 & 0 & 0 & \cdots & 0 & 0 \\
               0 & 0 & 0 & \cdots & 0 & 0 \\
               -1 & 0 & 0 & \cdots & 0 & 0
           \end{pmatrix}.
\end{array}}
\label{eq0217}
\end{equation}
Considering (\ref{eq0216}) for more consecutive time slots and using simple algebra, one may easily show that
\begin{equation}
\small{\bm{\bar{Y}}_j^l = \bm{H}_j\cdot\begin{pmatrix}
    \bm{S}_j^{l-2}, \bm{S}_j^{l-1}, \bm{S}_j^l \\
\end{pmatrix}\cdot\bm{\bar{A}} + \bm{\bar{N}}_j^l = \bm{H}_j \ \bm{S}_j^{l-2} \ \bm{U}_j^l \ \bm{\bar{A}} + \bm{\bar{N}}_j^l.}
\label{eq0218}
\end{equation}

\section{Proof of Theorem 5.1}
In the ITIC decoder using Method 0, we used the relationship in (\ref{eq0014}) and performed noncoherent ML detection. In (\ref{eq0014}), $\bm{H}_j \bm{S}_j^{l-2}$, $\bm{U}_j^l \bm{\bar{A}}$, and $\bm{\bar{N}}_j^l$ can be considered as the equivalent channel, signal, and noise terms, respectively. Note that the entries of $\bm{H}_j \bm{S}_j^{l-2}$ and $\bm{\bar{N}}_j^l$ are samples of independent zero-mean complex Gaussian random variables. With a small abuse of the notation, let $\bm{U}_{j,1}^l = (\bm{I}_N, \bm{P}_{j,1}^{l-1}, \bm{P}_{j,1}^{l-1}\bm{P}_{j,1}^l)$, $\bm{U}_{j,2}^l = (\bm{I}_N, \bm{P}_{j,2}^{l-1}, \bm{P}_{j,2}^{l-1}\bm{P}_{j,2}^l)$ for some arbitrary data matrices $\bm{P}_{j,1}^{l-1}, \bm{P}_{j,1}^l, \bm{P}_{j,2}^{l-1}, \bm{P}_{j,2}^l$ such that $\bm{U}_{j,1}^l \ne \bm{U}_{j,2}^l$. Then, in order to prove that the ITIC decoder using Method 0 achieves a diversity order of $MN$, by Proposition 4 of \cite{BV01}, it suffices to show that for any $\bm{U}_{j,1}^l \ne \bm{U}_{j,2}^l$, the following has full row rank\footnotemark:\footnotetext[2]{The channel model used in \cite{BV01} is the transposed version of ours. We have modified their results based on our channel model. We have also used the fact that $\rank(X^\dagger X)=\rank(X)$ for any matrix $X$ with complex elements.}
\begin{equation}
\begin{array}{l@{}l}
\begin{pmatrix}[0.75]
               \bm{U}_{j,1}^l \cdot \bm{\bar{A}} \\
               \bm{U}_{j,2}^l \cdot \bm{\bar{A}}
           \end{pmatrix} &= \begin{pmatrix}[0.75]
               \bm{I}_N \ \ \bm{P}_{j,1}^{l-1} \ \ \bm{P}_{j,1}^{l-1}\bm{P}_{j,1}^l \\
               \bm{I}_N \ \ \bm{P}_{j,2}^{l-1} \ \ \bm{P}_{j,2}^{l-1}\bm{P}_{j,2}^l
           \end{pmatrix} \cdot \bm{\bar{A}} \\
           &= \bm{G}(\bm{P}_{j,1}^{l-1}, \bm{P}_{j,1}^l, \bm{P}_{j,2}^{l-1}, \bm{P}_{j,2}^l).
\end{array}
\label{eq0339}
\end{equation}
By our assumption, $\bm{G}(\bm{P}_{j,1}^{l-1}, \bm{P}_{j,1}^l, \bm{P}_{j,2}^{l-1}, \bm{P}_{j,2}^l)$ has full row rank when $(\bm{P}_{j,1}^{l-1}, \bm{P}_{j,1}^l) \ne (\bm{P}_{j,2}^{l-1}, \bm{P}_{j,2}^l)$ (or equivalently, $\bm{U}_{j,1}^l \ne \bm{U}_{j,2}^l$). Thus, the ITIC decoder using Method 0 provides full diversity. Now, note that the MMPL decoder using Method 0 is optimal among the decoders using the same set of (or a subset of) the time partitions it uses. Since the ITIC decoder using Method 0 uses a subset of the time partitions the MMPL decoder using Method 0 uses, the MMPL decoder using Method 0 must perform at least as good as the ITIC decoder using Method 0. Thus, the MMPL decoder using Method 0 must achieve full diversity as well.

\section{Proof of Theorem 5.3}
We need the following property to prove the theorem:
\begin{lemma}
Let $\bm{X}_1, \bm{X}_2$ be distinct $N \times N$ matrices such that $(\bm{X}_2 - \bm{X}_1)^\dagger \cdot (\bm{X}_2 - \bm{X}_1) = \frac{\| \bm{X}_2 - \bm{X}_1 \|_F^2}{N} \cdot \bm{I}_N$. Then,
\begin{equation}
\footnotesize{\begin{pmatrix}[0.75]
               \bm{I}_N \ \ \bm{X}_1 \\
               \bm{I}_N \ \ \bm{X}_2
           \end{pmatrix}^{-1} = \begin{pmatrix}[0.75]
               \bm{I}_N + \bm{X}_1 \bm{\bar{X}} \ \ -\bm{X}_1 \bm{\bar{X}} \\
               -\bm{\bar{X}} \qquad \quad \bm{\bar{X}}
           \end{pmatrix}}
\label{eq0400}
\end{equation}
where $\bm{\bar{X}} = \frac{N(\bm{X}_2 - \bm{X}_1)^\dagger}{\| \bm{X}_2 - \bm{X}_1 \|_F^2}$.
\begin{proof}
The result can be easily proven by showing that
\begin{equation}
\footnotesize{\begin{pmatrix}[0.75]
               \bm{I}_N + \bm{X}_1 \bm{\bar{X}} \ \ -\bm{X}_1 \bm{\bar{X}} \\
               -\bm{\bar{X}} \qquad \quad \bm{\bar{X}}
           \end{pmatrix} \cdot \begin{pmatrix}[0.75]
               \bm{I}_N \ \ \bm{X}_1 \\
               \bm{I}_N \ \ \bm{X}_2
           \end{pmatrix} = \begin{pmatrix}[0.75]
               \bm{I}_N \ \ \bm{0}_N \\
               \bm{0}_N \ \ \bm{I}_N
           \end{pmatrix} = \bm{I}_{2N}.}
\label{eq0401}
\end{equation}
\end{proof}
\end{lemma}
To prove Theorem 5.3, we consider two cases:

\emph{Case 1:} We first consider the case when $\bm{P}_1 \ne \bm{P}_3$. Since $\bm{P}_1, \bm{P}_3$ are constructed using the same OSTBC and thus $(\bm{P}_3 -\nolinebreak \bm{P}_1)^\dagger \cdot (\bm{P}_3 - \bm{P}_1) = \frac{\| \bm{P}_3 - \bm{P}_1 \|_F^2}{N} \cdot \bm{I}_N$, by Lemma C.1, $\begin{pmatrix}[0.75] \bm{I}_N \ \ \bm{P}_1 \\ \bm{I}_N \ \ \bm{P}_3 \end{pmatrix}$ is invertible. Also, since its inverse must be a full rank matrix, multiplying its inverse by $\bm{G}(\bm{P}_1, \bm{P}_2, \bm{P}_3, \bm{P}_4)$ must result in a matrix with the same rank as $\bm{G}(\bm{P}_1, \bm{P}_2, \bm{P}_3, \bm{P}_4)$. Therefore, using Lemma C.1 and the definition of $\bm{G}(\bm{P}_1, \bm{P}_2, \bm{P}_3, \bm{P}_4)$ in (\ref{eq0314}), by multiplying $\bm{G}(\bm{P}_1, \bm{P}_2, \bm{P}_3, \bm{P}_4)$ by $\begin{pmatrix}[0.75] \bm{I}_N \ \ \bm{P}_1 \\ \bm{I}_N \ \ \bm{P}_3 \end{pmatrix}^{-1}$ from the left we obtain
\begin{equation}
\footnotesize{\setlength\arraycolsep{3pt}
\begin{array} {l}
\begin{pmatrix}[0.75] \bm{I}_N \ \ \bm{P}_1 \\ \bm{I}_N \ \ \bm{P}_3 \end{pmatrix}^{-1} \cdot \bm{G}(\bm{P}_1, \bm{P}_2, \bm{P}_3, \bm{P}_4) \\
= \begin{pmatrix}[0.75]
               \bm{I}_N + \bm{P}_1 \left(\frac{N(\bm{P}_3 - \bm{P}_1)^\dagger}{\| \bm{P}_3 - \bm{P}_1 \|_F^2}\right) \ \ -\bm{P}_1 \left(\frac{N(\bm{P}_3 - \bm{P}_1)^\dagger}{\| \bm{P}_3 - \bm{P}_1 \|_F^2}\right) \\
               -\frac{N(\bm{P}_3 - \bm{P}_1)^\dagger}{\| \bm{P}_3 - \bm{P}_1 \|_F^2} \qquad \quad \frac{N(\bm{P}_3 - \bm{P}_1)^\dagger}{\| \bm{P}_3 - \bm{P}_1 \|_F^2}
           \end{pmatrix} \\
\qquad \qquad \qquad \qquad \qquad \qquad \qquad \qquad \qquad \qquad \ \ \cdot \begin{pmatrix}[0.75]
               \bm{I}_N & \bm{P}_1 & \bm{P}_1 \bm{P}_2 \\
               \bm{I}_N & \bm{P}_3 & \bm{P}_3 \bm{P}_4
           \end{pmatrix} \cdot \bm{\bar{A}} \\
           = \begin{pmatrix}[0.75]
               \bm{I}_N & \bm{0}_N & \bm{P}_1\bm{P}_2 - N \bm{P}_1 \cdot \left(\frac{\left(\bm{P}_3 - \bm{P}_1\right)^\dagger}{\big\| \bm{P}_3 - \bm{P}_1 \big\|_F^2}\right) \cdot \left(\bm{P}_3\bm{P}_4 - \bm{P}_1\bm{P}_2\right) \\
               \bm{0}_N & \bm{I}_N & N \left(\frac{\left(\bm{P}_3 - \bm{P}_1\right)^\dagger}{\big\| \bm{P}_3 - \bm{P}_1 \big\|_F^2}\right) \cdot \left(\bm{P}_3\bm{P}_4 - \bm{P}_1\bm{P}_2\right)
           \end{pmatrix} \cdot \bm{\bar{A}},
\end{array}}
\label{eq0402}
\end{equation}
which must be of the same rank as $\bm{G}(\bm{P}_1, \bm{P}_2, \bm{P}_3, \bm{P}_4)$. Now, let $\bm{B}_1$ and $\bm{B}_1^{-1}$ be $3N-1 \times 3N-1$ matrices given by
\begin{equation}
\footnotesize{\begin{array} {c}
\bm{B}_1 = \begin{pmatrix}[0.75]
               1 & -1 & 0 & \cdots & 0 & 0 \\
               0 & 1 & -1 & \cdots & 0 & 0 \\
               0 & 0 & 1 & \cdots & 0 & 0 \\
               \ddots & \ddots & \ddots & \ddots & \ddots & \ddots \\
               0 & 0 & 0 & \cdots & -1 & 0 \\
               0 & 0 & 0 & \cdots & 1 & -1 \\
               0 & 0 & 0 & \cdots & 0 & 1
           \end{pmatrix}, \\\\
\bm{B}_1^{-1} = \begin{pmatrix}[0.75]
               1 & 1 & 1 & \cdots & 1 & 1 \\
               0 & 1 & 1 & \cdots & 1 & 1 \\
               0 & 0 & 1 & \cdots & 1 & 1 \\
               \ddots & \ddots & \ddots & \ddots & \ddots & \ddots \\
               0 & 0 & 0 & \cdots & 1 & 1 \\
               0 & 0 & 0 & \cdots & 1 & 1 \\
               0 & 0 & 0 & \cdots & 0 & 1
           \end{pmatrix}.
\end{array}}
\label{eq0342}
\end{equation}
Note that $\bm{B}_1^{-1}$ is the inverse of $\bm{B}_1$. Again, since $\bm{B}_1^{-1}$ is a full rank matrix, multiplying it by (\ref{eq0402}) will result in a matrix with the same rank as (\ref{eq0402}). Therefore, multiplying (\ref{eq0402}) by $\bm{B}_1^{-1}$ from the right yields a matrix with the same rank as $\bm{G}(\bm{P}_1, \bm{P}_2, \bm{P}_3, \bm{P}_4)$, given by
\begin{equation}
\footnotesize{\setlength\arraycolsep{3pt}
\begin{array} {l}
\begin{pmatrix}[0.75] \bm{I}_N \ \ \bm{P}_1 \\ \bm{I}_N \ \ \bm{P}_3 \end{pmatrix}^{-1} \cdot \bm{G}(\bm{P}_1, \bm{P}_2, \bm{P}_3, \bm{P}_4) \cdot \bm{B}_1^{-1} \\
= \begin{pmatrix}[0.75]
               \bm{I}_N & \bm{0}_N & \bm{P}_1\bm{P}_2 - N \bm{P}_1 \cdot \left(\frac{\left(\bm{P}_3 - \bm{P}_1\right)^\dagger}{\big\| \bm{P}_3 - \bm{P}_1 \big\|_F^2}\right) \cdot \left(\bm{P}_3\bm{P}_4 - \bm{P}_1\bm{P}_2\right) \\
               \bm{0}_N & \bm{I}_N & N \left(\frac{\left(\bm{P}_3 - \bm{P}_1\right)^\dagger}{\big\| \bm{P}_3 - \bm{P}_1 \big\|_F^2}\right) \cdot \left(\bm{P}_3\bm{P}_4 - \bm{P}_1\bm{P}_2\right)
           \end{pmatrix} \cdot \bm{B}_2
\end{array}}
\label{eq0403}
\end{equation}
where $\bm{B}_2$ is the $3N \times 3N-1$ matrix
\begin{equation}
\footnotesize{\bm{B}_2 = \bm{\bar{A}} \cdot \bm{B}_1^{-1} = \begin{pmatrix}[0.75]
               -1 & -1 & -1 & \cdots & -1 & -1 & -1 \\
               1 & 0 & 0 & \cdots & 0 & 0 & 0 \\
               0 & 1 & 0 & \cdots & 0 & 0 & 0 \\
               0 & 0 & 1 & \cdots & 0 & 0 & 0 \\
               \ddots & \ddots & \ddots & \ddots & \ddots & \ddots & \ddots \\
               0 & 0 & 0 & \cdots & 1 & 0 & 0 \\
               0 & 0 & 0 & \cdots & 0 & 1 & 0 \\
               0 & 0 & 0 & \cdots & 0 & 0 & 1
           \end{pmatrix}.}
\label{eq0404}
\end{equation}
Now, consider the RHS of (\ref{eq0403}) and let
\begin{equation}
\small{\begin{array} {l}
\begin{pmatrix}[0.75]
               \bm{P}_1\bm{P}_2 - N \bm{P}_1 \cdot \left(\frac{\left(\bm{P}_3 - \bm{P}_1\right)^\dagger}{\big\| \bm{P}_3 - \bm{P}_1 \big\|_F^2}\right) \cdot \left(\bm{P}_3\bm{P}_4 - \bm{P}_1\bm{P}_2\right) \\
               N \left(\frac{\left(\bm{P}_3 - \bm{P}_1\right)^\dagger}{\big\| \bm{P}_3 - \bm{P}_1 \big\|_F^2}\right) \cdot \left(\bm{P}_3\bm{P}_4 - \bm{P}_1\bm{P}_2\right)
           \end{pmatrix} \\
\qquad \qquad \qquad \qquad \ \ = \begin{pmatrix}[0.75]
               \beta_{1,1} & \beta_{1,2} & \beta_{1,3} & \cdots & \beta_{1,N} \\
               \beta_{2,1} & \beta_{2,2} & \beta_{2,3} & \cdots & \beta_{2,N} \\
               \beta_{3,1} & \beta_{3,2} & \beta_{3,3} & \cdots & \beta_{3,N} \\
               \ddots & \ddots & \ddots & \ddots & \ddots \\
               \beta_{2N,1} & \beta_{2N,2} & \beta_{2N,3} & \cdots & \beta_{2N,N}
           \end{pmatrix}.
\end{array}}
\label{eq0405}
\end{equation}
By plugging (\ref{eq0405}) into (\ref{eq0403}) and using simple algebra, we can write (\ref{eq0403}) as
\begin{equation}
\footnotesize{\setlength\arraycolsep{2pt}
\begin{array} {l}
\begin{pmatrix}[0.75] \bm{I}_N \ \ \bm{P}_1 \\ \bm{I}_N \ \ \bm{P}_3 \end{pmatrix}^{-1} \cdot \bm{G}(\bm{P}_1, \bm{P}_2, \bm{P}_3, \bm{P}_4) \cdot \bm{B}_1^{-1} = \\\\\\
\begin{pmatrix}[0.75]
               \smash{\overbrace{-1 \ -1 \ -1 \quad \cdots \ -1}^{2N-1\ \textrm{times}}} & \beta_{1,1}-1 & \beta_{1,2}-1 & \beta_{1,3}-1 & \cdots & \beta_{1,N}-1 \\
               1 \ \ \quad 0 \ \ \quad 0 \quad \cdots \quad 0 & \beta_{2,1} & \beta_{2,2} & \beta_{2,3} & \cdots & \beta_{2,N} \\
               0 \ \ \quad 1 \ \ \quad 0 \quad \cdots \quad 0 & \beta_{3,1} & \beta_{3,2} & \beta_{3,3} & \cdots & \beta_{3,N} \\
               0 \ \ \quad 0 \ \ \quad 1 \quad \cdots \quad 0 & \beta_{4,1} & \beta_{4,2} & \beta_{4,3} & \cdots & \beta_{4,N} \\
               \ddots \quad \ddots \quad \ddots \quad \ddots \quad \ddots & \ddots & \ddots & \ddots & \ddots & \ddots \\
               0 \ \ \quad 0 \ \ \quad 0 \quad \cdots \quad 1 & \beta_{2N,1} & \beta_{2N,2} & \beta_{2N,3} & \cdots & \beta_{2N,N}
           \end{pmatrix}.
\end{array}}
\label{eq0406}
\end{equation}
Let $\bm{r}_i$, $i=1,\cdots,2N$, denote the $i$th row of (\ref{eq0406}). Then, the linear combination of $\bm{r}_1, \cdots, \bm{r}_{2N}$ with coefficients $\lambda_1, \lambda_2, \cdots, \lambda_{2N}$, which are not all zero, is given by
\begin{equation}
\scriptsize{\begin{array}{l@{}l}
\bm{r} &= \displaystyle\sum_{i=1}^{2N} \lambda_i\bm{r}_i \\
&= \left(\lambda_2-\lambda_1, \cdots, \lambda_{2N}-\lambda_1, -\lambda_1+\displaystyle\sum_{i=1}^{2N} \lambda_i\beta_{i,1}, \cdots, -\lambda_1+\displaystyle\sum_{i=1}^{2N} \lambda_i\beta_{i,N}\right).
\end{array}}
\label{eq0407}
\end{equation}
Note that $\bm{r}$ is equal to the zero vector if and only if $\lambda_1 = \lambda_2 = \cdots = \lambda_{2N}$ and $\sum_{i=1}^{2N} \beta_{i,1} = \sum_{i=1}^{2N} \beta_{i,2} = \cdots = \sum_{i=1}^{2N} \beta_{i,N} = 1$. This means that the rows of (\ref{eq0406}) are linearly dependent if and only if $\sum_{i=1}^{2N} \beta_{i,1} = \sum_{i=1}^{2N} \beta_{i,2} = \cdots = \sum_{i=1}^{2N} \beta_{i,N} = 1$. Using (\ref{eq0405}), this implies that (\ref{eq0406}), and thus $\bm{G}(\bm{P}_1, \bm{P}_2, \bm{P}_3, \bm{P}_4)$, has full row rank if and only if
\begin{equation}
\tiny{\left(\smash{\overbrace{1,1,\cdots,1}^{2N \textrm{times}}}\right) \cdot \begin{pmatrix}[0.75]
               \bm{P}_1\bm{P}_2 - N \bm{P}_1 \cdot \left(\frac{\left(\bm{P}_3 - \bm{P}_1\right)^\dagger}{\big\| \bm{P}_3 - \bm{P}_1 \big\|_F^2}\right) \cdot \left(\bm{P}_3\bm{P}_4 - \bm{P}_1\bm{P}_2\right) \\
               N \left(\frac{\left(\bm{P}_3 - \bm{P}_1\right)^\dagger}{\big\| \bm{P}_3 - \bm{P}_1 \big\|_F^2}\right) \cdot \left(\bm{P}_3\bm{P}_4 - \bm{P}_1\bm{P}_2\right)
           \end{pmatrix} \ne \left(\smash{\overbrace{1,1,\cdots,1}^{N \textrm{times}}}\right).}
\label{eq0408}
\end{equation}
Then, it is easy to see that (\ref{eq0408}) holds, and thus $\bm{G}(\bm{P}_1, \bm{P}_2, \bm{P}_3, \bm{P}_4)$ has full row rank, for any possible data matrices $\bm{P}_1, \bm{P}_2, \bm{P}_3, \bm{P}_4$ with $\bm{P}_1 \ne \bm{P}_3$ if and only if (\ref{eq0340}) holds for any possible data matrices $\bm{\tilde{P}}_1, \bm{\tilde{P}}_2, \bm{\tilde{P}}_3, \bm{\tilde{P}}_4$ with $\bm{\tilde{P}}_1 \ne \bm{\tilde{P}}_3$. This means that (\ref{eq0340}) is a necessary and sufficient condition for $\bm{G}(\bm{P}_1, \bm{P}_2, \bm{P}_3, \bm{P}_4)$ to have full row rank in Case 1.

\emph{Case 2:} We now consider the case when $\bm{P}_1 = \bm{P}_3$. Since $(\bm{P}_1, \bm{P}_2) \ne (\bm{P}_3, \bm{P}_4)$, this implies that $\bm{P}_2 \ne \bm{P}_4$. Also, since $\bm{P}_2, \bm{P}_4$ are constructed using the same OSTBC and thus $[\bm{P}_1(\bm{P}_4 - \bm{P}_2)]^\dagger \cdot [\bm{P}_1(\bm{P}_4 - \bm{P}_2)] = \frac{\| \bm{P}_1(\bm{P}_4 - \bm{P}_2) \|_F^2}{N} \cdot \bm{I}_N$, by Lemma C.1, $\begin{pmatrix}[0.75] \bm{I}_N \ \ \bm{P}_1\bm{P}_2 \\ \bm{I}_N \ \ \bm{P}_1\bm{P}_4 \end{pmatrix}$ is invertible. Again, since its inverse must be a full rank matrix, multiplying its inverse by $\bm{G}(\bm{P}_1, \bm{P}_2, \bm{P}_3, \bm{P}_4)$ must result in a matrix with the same rank as $\bm{G}(\bm{P}_1, \bm{P}_2, \bm{P}_3, \bm{P}_4)$. Therefore, by multiplying $\bm{G}(\bm{P}_1, \bm{P}_2, \bm{P}_3, \bm{P}_4)$ by $\begin{pmatrix}[0.75] \bm{I}_N \ \ \bm{P}_1\bm{P}_2 \\ \bm{I}_N \ \ \bm{P}_1\bm{P}_4 \end{pmatrix}^{-1}$ from the left we obtain
\begin{equation}
\footnotesize{\begin{array} {l}
\begin{pmatrix}[0.75] \bm{I}_N \ \ \bm{P}_1\bm{P}_2 \\ \bm{I}_N \ \ \bm{P}_1\bm{P}_4 \end{pmatrix}^{-1} \cdot \bm{G}(\bm{P}_1, \bm{P}_2, \bm{P}_3, \bm{P}_4) \\
= \begin{pmatrix}[0.75]
               \bm{I}_N + \bm{P}_1\bm{P}_2\left(\frac{N[\bm{P}_1(\bm{P}_4 - \bm{P}_2)]^\dagger}{\| \bm{P}_1(\bm{P}_4 - \bm{P}_2) \|_F^2}\right) \ \ - \bm{P}_1\bm{P}_2\left(\frac{N[\bm{P}_1(\bm{P}_4 - \bm{P}_2)]^\dagger}{\| \bm{P}_1(\bm{P}_4 - \bm{P}_2) \|_F^2}\right) \\
               -\frac{N[\bm{P}_1(\bm{P}_4 - \bm{P}_2)]^\dagger}{\| \bm{P}_1(\bm{P}_4 - \bm{P}_2) \|_F^2} \qquad \qquad \frac{N[\bm{P}_1(\bm{P}_4 - \bm{P}_2)]^\dagger}{\| \bm{P}_1(\bm{P}_4 - \bm{P}_2) \|_F^2}
           \end{pmatrix} \\
\qquad \qquad \qquad \qquad \qquad \qquad \qquad \qquad \qquad \quad \ \ \cdot \begin{pmatrix}[0.75]
               \bm{I}_N & \bm{P}_1 & \bm{P}_1 \bm{P}_2 \\
               \bm{I}_N & \bm{P}_1 & \bm{P}_1 \bm{P}_4
           \end{pmatrix} \cdot \bm{\bar{A}} \\
= \begin{pmatrix}[0.75]
               \bm{I}_N & \bm{P}_1 & \bm{0}_N \\
               \bm{0}_N & \bm{0}_N & \bm{I}_N
           \end{pmatrix} \cdot \bm{\bar{A}},
\end{array}}
\label{eq0409}
\end{equation}
which must be of the same rank as $\bm{G}(\bm{P}_1, \bm{P}_2, \bm{P}_3, \bm{P}_4)$. Once again, since $\bm{B}_1^{-1}$ is a full rank matrix, multiplying (\ref{eq0409}) by $\bm{B}_1^{-1}$ from the right yields a matrix with the same rank as (\ref{eq0409}), and thus $\bm{G}(\bm{P}_1, \bm{P}_2, \bm{P}_3, \bm{P}_4)$, given by
\begin{equation}
\footnotesize{\begin{pmatrix}[0.75] \bm{I}_N \ \ \bm{P}_1\bm{P}_2 \\ \bm{I}_N \ \ \bm{P}_1\bm{P}_4 \end{pmatrix}^{-1} \cdot \bm{G}(\bm{P}_1, \bm{P}_2, \bm{P}_3, \bm{P}_4) \cdot \bm{B}_1^{-1} = \begin{pmatrix}[0.75]
               \bm{I}_N & \bm{P}_1 & \bm{0}_N \\
               \bm{0}_N & \bm{0}_N & \bm{I}_N
           \end{pmatrix} \cdot \bm{B}_2.}
\label{eq0410}
\end{equation}
Then proceeding similarly to the procedure described in (\ref{eq0405})-(\ref{eq0408}) for Case 1, we find that $\bm{G}(\bm{P}_1, \bm{P}_2, \bm{P}_3, \bm{P}_4)$ has full row rank if and only if $\bm{w} \cdot \bm{P}_1 \ne \bm{w}$. Note that this condition is a special case of (\ref{eq0340}) when $\bm{\tilde{P}}_2 = \bm{\tilde{P}}_4 = \bm{P}_1$. Therefore, (\ref{eq0340}) is a sufficient condition for $\bm{G}(\bm{P}_1, \bm{P}_2, \bm{P}_3, \bm{P}_4)$ to have full row rank in Case 2. Also, we showed that (\ref{eq0340}) is a necessary and sufficient condition for $\bm{G}(\bm{P}_1, \bm{P}_2, \bm{P}_3, \bm{P}_4)$ to have full row rank in Case 1. Thus, (\ref{eq0340}) is a necessary and sufficient condition in the general case for $\bm{G}(\bm{P}_1, \bm{P}_2, \bm{P}_3, \bm{P}_4)$ to have full row rank for any possible data matrices $\bm{P}_1, \bm{P}_2, \bm{P}_3, \bm{P}_4$ with $(\bm{P}_1, \bm{P}_2) \ne (\bm{P}_3, \bm{P}_4)$.

\ifCLASSOPTIONcaptionsoff
  \newpage
\fi



%
%

%

\begin{IEEEbiography}[{\includegraphics[width=1in,height=1.25in,clip,keepaspectratio]{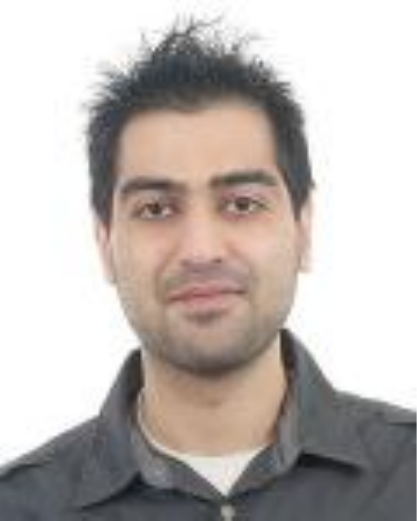}}]{Sina Poorkasmaei} received the B.S. degree in electrical and computer engineering from the American University in Dubai, UAE, in 2007, and the M.S. degree in electrical engineering and computer science from the University of California, Irvine, CA, USA, in 2008, where he is currently working toward the Ph.D. degree. His research interests include wireless communications and signal processing.
\end{IEEEbiography}
\begin{IEEEbiography}[{\includegraphics[width=1in,height=1.25in,clip,keepaspectratio]{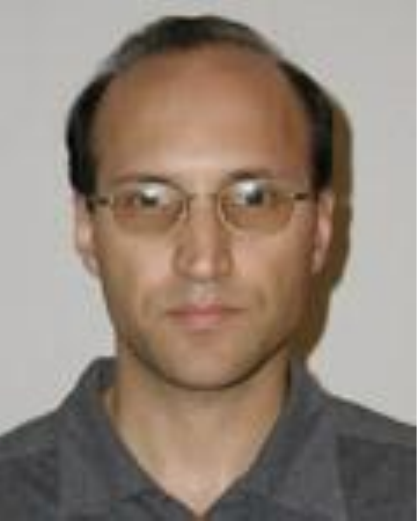}}]{Hamid Jafarkhani} (F'06) is a Chancellor's Professor with the Department of Electrical Engineering and Computer Science, University of California, where he is also the Director of the Center for Pervasive Communications and Computing and the Conexant-Broadcom Endowed Chair. He is the author of the book \emph{Space-Time Coding: Theory and Practice}.

Dr. Jafarkhani is a Fellow of AAAS. In 1984, he ranked first in the nationwide entrance examination of Iranian universities. He was the corecipient of the American Division Award of the 1995 Texas Instruments DSP Solutions Challenge. He was the recipient of an NSF Career Award in 2003. He was the recipient of the UCI Distinguished Mid-Career Faculty Award for Research in 2006 and the School of Engineering Fariborz Maseeh Best Faculty Research Award in 2007. In addition, he was the corecipient of the 2002 Best Paper Award of ISWC, the 2006 IEEE Marconi Best Paper Award in Wireless Communications, the 2009 Best Paper Award of the JOURNAL OF COMMUNICATIONS AND NETWORKS, the 2012 IEEE Globecom Best Paper Award (Communication Theory Symposium), the 2013 IEEE Eric E. Sumner Award, and the 2014 IEEE Communications Society Award for Advances in Communication. He is listed as a highly cited researcher in http://www.isihighlycited.com. During 1997-2007, according to the Thomson Scientific, he was one of the top 10 most cited researchers in the field of ``computer science.''
%
%
\end{IEEEbiography}

\end{document}